\documentclass[a4paper,aps,11pt,nofootinbib]{revtex4}
\usepackage{amsmath}
\usepackage{amssymb}
\usepackage{amsfonts}
\usepackage[dvipdfmx]{graphicx}
\usepackage{color}
\usepackage{float}
\usepackage{comment}
\usepackage{ulem}
\usepackage{ascmac}
\usepackage{subfigure} 
\usepackage{setspace}

\definecolor{DarkBlue}{rgb}{0,0,0.7}

\definecolor{DarkRed}{rgb}{0.65,0,0}

\definecolor{DarkGreen}{rgb}{0,0.3,0}

\definecolor{purple}{rgb}{0.7,0,0.7}

\begin{document}

\def\uwave{ }
\title{\Large
Relativistic Nontopological Soliton Stars \\
in a U(1) Gauge Higgs Model}
\today

\hfill{OCU-PHYS 558}

\hfill{AP-GR 179}

\hfill{NITEP 132}

\author{\vspace{1cm}
\Large Yota Endo}
\email{yota-endo@dt.osaka-cu.ac.jp}
\author{
\Large Hideki Ishihara}
\email{ishihara@osaka-cu.ac.jp}
\author{\Large Tatsuya Ogawa}
\email{taogawa@osaka-cu.ac.jp}
\affiliation{
 Department of Mathematics and Physics,
 Graduate School of Science,
Nambu Yoichiro Institute of Theoretical and Experimental Physics (NITEP),
Osaka City University, Osaka 558-8585, Japan}

\bigskip

\begin{abstract}
\bigskip

We study spherically symmetric nontopological soliton stars (NTS stars) numerically
in the coupled system of a complex scalar field, a U(1) gauge field,
a complex Higgs scalar field, and Einstein gravity,
where the symmetry is broken spontaneously.
The gravitational mass of NTS stars is limited by a maximum mass for a fixed breaking scale,
and the maximum mass increases steeply as the breaking scale decreases.
In the case of the breaking scale is much less than the Planck scale,
the maximum mass of NTS stars becomes the astrophysical scale, and such a star is
relativistically compact so that it has the innermost stable circular orbit.

The first author contributed with a part of numerical calculations.
The second did with planning and conducting the research, and
the third did with all numerical calculations and finding
new properties of the system.
\end{abstract}

\maketitle

\newpage
\section{Introduction}

Nontopological solitons (NTSs) are localized solutions carrying a Noether charge
in nonlinear field theories that have a continuous global symmetry.
Rosen \cite{Rosen:1968mfz}, in his pioneering work, showed that a self-interacting
complex scalar field theory admits particlelike NTS solutions,
and Coleman \cite{Coleman:1985ki}
proved the existence theorem of spherically symmetric NTS solutions with a conserved charge,
he called them \lq Q-balls\rq\ ,
in nonlinearly self-coupling complex scalar field theories with some
conditions.
Friedberg, Lee, and Sirlin \cite{Friedberg:1976me} studied a coupled system of a complex scalar field and
a real scalar field with a double-well potential, and showed the existence of
the NTS solutions
(see e.g. reviews \cite{Lee:1991ax,Nugaev:2019vru} and textbooks \cite{Shnir:2018yzp}).
The NTS solutions in extended field theories including a gauge field were also
studied in the works \cite{Lee:1988ag,Shi:1991gh, Gulamov:2015fya}.
The NTSs are interested as a possible candidate of
dark matter \cite{Kusenko:1997si, Kusenko:2001vu, Fujii:2001xp,Enqvist:2001jd,
Kusenko:2004yw},
and as sources for baryogenesis \cite{Enqvist:1997si,Kasuya:1999wu,Kawasaki:2002hq}.

Recently, NTS solutions were constructed in the theory that consists of a complex scalar field,
a U(1) gauge field, and a complex Higgs scalar field with a Mexican hat potential
which causes the spontaneous symmetery breaking \cite{Ishihara:2018rxg,Ishihara:2019gim,Ishihara:2021iag}
\footnote{
NTSs in a generalized model are discussed in Refs.\cite{Forgacs:2020vcy, Forgacs:2020sms}.
}.
In this model, there are interesting properties:
the charges carried by two scalar fields are screening each other;
and NTS solutions with infinitely large mass can exist \cite{Ishihara:2021iag}.
These would suggest that NTSs with astrophysical scale in this model
can play important roles in cosmology.
However, infinite mass should be prohibited if we take gravity produced
by the NTS into account,
namely, the mass should be limited by a maximum mass for self-gravitaing NTSs.

It is also investigated that
localized objects made by self-gravitating complex scalar fields, so-called {\it boson stars}.
In the model of a free massive complex scalar field with gravity,
the gravitational mass of the localized solutions is quite small
then the solutions are called {\it mini boson stars} \cite{Kaup:1968zz, Ruffini:1969qy},
while if the complex scalar field has nonlinear self-coupling,
the mass of solution can be
large \cite{Colpi:1986ye}.
Boson stars in various field models are studied in Refs.\cite{Jetzer:1989av, Henriques:1989ar,Henriques:1989ez}
(see also \cite{Jetzer:1991jr,Schunck:2003kk,Liebling:2012fv} for review).
Furthermore, self-gravity of the NTS
is also studied
in the Colman's model \cite{Lynn:1988rb} and
the Friedberg-Lee-Sirlin's model \cite{Lee:1986ts,Friedberg:1986tq,Lee:1986tr, Kunz:2021mbm}.

In this paper, we consider the coupled system of two complex scalar fields
and a U(1) gauge field, which is studied in Refs.\cite{Ishihara:2018rxg,Ishihara:2019gim,Ishihara:2021iag},
with Einstein gravity.
The local U(1) symmetry of the system is spontaneously broken in a vacuum state where
one of the scalar field has an expectation value.
The system has two dimensionful parameters, the symmetry breaking scale, $\eta$,
and the Plank scale, $M_\text{P}$,
then the dimensionless parameter $\eta/M_\text{P}$ is an important quantity
that characterizes the model.

Assuming spherically symmetry, stationary rotation of the phase of complex scalar fields,
and static geometry, we derive a set of coupled ordinary differential equations.
We obtain numerical solutions that describe
self-gravitating objects, we call them \lq nontopological soliton stars (NTS stars)\rq\
in this paper,
and investigate properties of the solutions, especially, mass and radius
that depend on $\eta/M_\text{P}$.
It is interesting that NTS stars with mass of astrophysical scale are possible,
e.g., the solar mass is possible for $\eta\sim $1GeV,
and the NTS stars can be so compact that they have the innermost stable circular orbits for $\eta/M_\text{P}\ll 1$.

The paper is organized as follows.
In Sec.II, we present the model that has a symmetry breaking scale,
and derive basic equations on the assumptions of the geometrical symmetries of the fields.
In Sec.III, we solve the basic equations numerically, and present NTS star solutions.
In Sec.IV, we investigate internal structures of the NTS stars:
energy density, pressure, and charge density.
In Sec.V, we study
mass and radius of the NTS stars, and see that
the maximum mass appears for each breaking scale.
Paying attention to the NTS stars with maximum mass in various breaking scales,
we investigate scale dependence of the maximum mass and the compactness in Sec.VI.
Section VII is devoted to conclusions.

\section{Basic Model}
We consider the action
\begin{align}
	S=\int \sqrt{-g}d^4x \left(\frac{R}{16\pi G}+\mathcal{L}_m\right),
\label{eq:action}
\end{align}
where $R$ is the scalar curvature with respect to a metric, $g_{\mu\nu}$, $g:={\rm det}(g_{\mu\nu})$, and $G$ is the gravitational constant.
The matter Lagrangian, $\mathcal{L}_m$, is given by
\begin{align}
  \mathcal{L}_m=& -g^{\mu\nu}(D_{\mu}\psi)^{\ast}(D_{\nu}\psi)-g^{\mu\nu}(D_{\mu}\phi)^{\ast}(D_{\nu}\phi)
	-\frac{\lambda}{4}(\left|\phi \right|^2-\eta^2)^2
\cr
	&-\mu \left|\phi \right|^2\left|\psi \right|^2
	-\frac{1}{4}g^{\mu\alpha}g^{\nu\beta}F_{\mu\nu}F_{\alpha\beta},
\label{eq:lagrangian}
\end{align}
where it consists of a complex matter scalar field $\psi$,
the field strength $F_{\mu\nu}:=\partial_{\mu}A_{\nu}-\partial_{\nu}A_{\mu}$
of a U(1) gauge field $A_{\mu}$,
and a complex Higgs scalar field $\phi$.
The gauge field couples to the scalar fields through the gauge covariant derivative,
$D_{\mu}:=\partial_{\mu}-ieA_{\mu}$.
The parameter $e$ is the charge of the fields $\psi$ and $\phi$, $\lambda$ the Higgs-self coupling, $\mu$ the Higgs-scalar coupling, and $\eta$ the Higgs vacuum expectation value determining the scale of the symmetry breaking.

The Lagrangian \eqref{eq:lagrangian} has local U(1) $\times$ global U(1) symmetry under the gauge
transformation given by
\begin{align}
  	&\psi(x) \to \tilde\psi(x)=e^{i(\chi(x)-\gamma)}\psi(x),
 \label{eq:psi_tr} \\
  	&\phi(x) \to \tilde\phi(x)=e^{i(\chi(x)+\gamma)}\phi(x),
 \label{eq:phi_tr} \\
 	&A_{\mu}(x)\to \tilde A_{\mu}(x)=A_{\mu}(x)+e^{-1}\partial_{\mu}\chi(x),
 \label{eq:A_tr}
\end{align}
where $\chi(x)$ is an arbitrary function that depends on spacetime coordinate, and $\gamma$ is
an arbitrary constant.
Concering to this invariance, conserved currents,
\begin{align}
j^{\mu}_{\psi}:=ie\left(\psi^{\ast}(D^{\mu}\psi)-(D^{\mu}\psi)^{\ast}\psi\right), \quad
j^{\mu}_{\phi}:=ie\left(\phi^{\ast}(D^{\mu}\phi)-(D^{\mu}\phi)^{\ast}\phi\right)
  \label{eq:current_phi_psi}
\end{align}
and conserved charges
\begin{align}
Q_{\psi}:=\int d^3x \sqrt{-g}\rho_{\psi} , \quad
Q_{\phi}:=\int d^3x \sqrt{-g}\rho_{\phi},
  \label{eq:charge_phi_psi}
\end{align}
are defined, where $\rho_{\psi}:=j^t_{\psi}$ and $\rho_{\phi}:=j^t_{\phi}$ are charge densities induced by the complex scalar field $\psi$ and $\phi$, respectively.
The integrations in \eqref{eq:charge_phi_psi} are performed on a timeslice $t=\text{\it const.}$

Owing to the potential term of the complex Higgs scalar field $\phi$ in \eqref{eq:lagrangian},
$\phi$ takes a nonzero vacuum expectation value $\eta$ in a vacuum state.
As a result, the symmetry is spontaneously broken, then, the scalar field $\psi$
and the U(1) gauge field $A_{\mu}$ acquire masses, $m_{\psi}:=\sqrt{\mu}\eta$
and $m_{A}:=\sqrt{2}e\eta$, respectively, through interactions with the complex Higgs field.
Simultaneously, a real scalar field as a fluctuation of the amplitude of $\phi$
around $\eta$ also acquires the mass $m_\phi:=\sqrt{\lambda}\eta$.

From the action \eqref{eq:action}, we can derive field equations
\begin{align}
  &\frac{1}{\sqrt{-g}}D_{\mu}\left(\sqrt{-g}g^{\mu\nu}D_{\nu}\psi\right)-\mu \psi \left|\phi \right|^2=0,
  \label{eq:ELequation_psi}
  \\
  &\frac{1}{\sqrt{-g}}D_{\mu}\left(\sqrt{-g}g^{\mu\nu}D_{\nu}\phi\right)-\frac{\lambda}{2}\phi(\left|\phi \right|^2-\eta^2)-\mu \left|\psi \right|^2 \phi=0,
 \label{eq:ELequation_phi}
 \\
 &\frac{1}{\sqrt{-g}}\partial_{\mu}(\sqrt{-g}F^{\mu\nu})=j^{\nu}_{\psi}+j^{\nu}_{\phi},
\label{eq:ELequation_A}
\\
 &G_{\mu\nu}=8\pi GT_{\mu\nu},
\label{eq:Einsteinequation}
\end{align}
where $G_{\mu\nu}:=R_{\mu\nu}-\frac{1}{2}g_{\mu\nu}R$ is the Einstein tensor and $T_{\mu\nu}$ is the energy-momentum tensor given by
\begin{align}
T_{\mu\nu}:=&-\frac{2}{\sqrt{-g}}\frac{\delta(\sqrt{-g}\mathcal{L}_m)}{\delta g^{\mu\nu}}
\notag \\
	=&2(D_{\mu}\psi)^{\ast}(D_{\nu}\psi)
	-g_{\mu\nu}
		(D_{\alpha}\psi)^{\ast}(D^{\alpha}\psi)
	+2(D_{\mu}\phi)^{\ast}(D_{\nu}\phi)
	-g_{\mu\nu}(D_{\alpha}\phi)^{\ast}(D^{\alpha}\phi)
\cr
	&-g_{\mu\nu}\left( \frac{\lambda}{4}(|\phi|^2-\eta^2)^2
	 + \mu |\psi|^2|\psi|^2 \right)
	+\left( F_{\mu\alpha}F_{\nu}^{~\alpha}
	-\frac{1}{4}g_{\mu\nu}F_{\alpha\beta}F^{\alpha\beta}\right).
\label{eq:T_munu}
\end{align}

Here, we assume stationary and spherically symmetric fields in the form:
\begin{align}
  	\psi=e^{-i\omega t}u(r), \quad \phi=e^{-i\omega' t}f(r), \quad
	A=A_t(r)dt ,
  \label{eq:matter_ansatz}
\end{align}
where we use a spherical coordinate $(t,r,\theta,\varphi)$.
The parameters $\omega$ and $\omega'$ are constant angular frequencies of the complex scalar fields.
Owing to the gauge transformation \eqref{eq:psi_tr}-\eqref{eq:A_tr}, we can fix the variables as
\begin{align}
	\phi(r) \to f(r), \quad \psi(t,r) \to e^{i\Omega t}u(r), \quad
	A_t(r) \to \alpha(r):= A_t(r)+e^{-1}\omega',
\label{eq:matter_variable}
\end{align}
where $\Omega:=\omega'-\omega$ is the parameter which characterizes the solution,
and takes a positive value without loss of generality.

We also take static and spherically symmetric spacetime assumptions in the form
\begin{align}
  ds^2=&g_{\mu\nu}dx^{\mu}dx^{\nu}\notag \\
  =&-\sigma(r)^2\left(1-\frac{2m(r)}{r}\right)dt^2+\left(1-\frac{2m(r)}{r}\right)^{-1}dr^2+r^2d\theta^2+r^2\sin^2\theta d\varphi^2,
 \label{eq:metric}
\end{align}
where $\sigma(r)$ and $m(r)$ are functions of $r$.
Note that $\sigma(r)$ is dimensionless, and $m(r)$ has dimension of length.
The Einstein equations reduce to
\begin{align}
	G^t_t=8\pi GT^t_t&, \quad G^r_r=8\pi GT^r_r, \quad
	G^{\theta}_{\theta}=G^{\varphi}_{\varphi}
		=8\pi GT^{\theta}_{\theta}=8\pi GT^{\varphi}_{\varphi}.
 \label{eq:Einsteinequation2}
\end{align}

Substituting assumptions \eqref{eq:matter_variable} and \eqref{eq:metric}
into \eqref{eq:ELequation_psi}$-$\eqref{eq:ELequation_A},
we obtain equations for the fields $\psi,\phi,$ and $A$ to be solved in the form:
\begin{align}
  &u''+\left(\frac{2}{r}\left(1+\frac{m-rm'}{r-2m}\right)+\frac{\sigma'}{\sigma}\right)u'
+\left(1-\frac{2m}{r}\right)^{-1}\left(\frac{(e\alpha-\Omega)^2u}{\sigma^2(1-2m/r)}-\mu f^2u\right)=0,
  \label{eq:equation_u}
  \\
 &f''+\left(\frac{2}{r}\left(1+\frac{m-rm'}{r-2m}\right)+\frac{\sigma'}{\sigma}\right)f'
\cr
	&\hspace{2cm} +\left(1-\frac{2m}{r}\right)^{-1}\left(\frac{e^2f\alpha^2}{\sigma^2(1-2m/r)}
	-\frac{\lambda}{2}f(f^2-1)	-\mu fu^2\right)=0,
 \label{eq:equation_f}
 \\
  &\alpha''+\left(\frac{2}{r}-\frac{\sigma'}{\sigma}\right)\alpha'
	+\left(1-\frac{2m}{r}\right)^{-1}\left(-2e^2f^2\alpha-2e(e\alpha-\Omega) u^2\right)=0,
 \label{eq:equation_alpha}
 \end{align}
where the prime denotes the derivative with respect to $r$.
Here and hereafter, $r$, $u(r)$, $f(r)$, $\alpha(r)$, $m(r)$, and $\Omega$ are normalized by $\eta$.


As for the Einstein equations,
we solve the time-time component of \eqref{eq:Einsteinequation2} and the combination
\begin{align}
	G^r_r-G^t_t=8\pi G(T^r_r-T^t_t).
 \label{eq:Einsteinequation3}
\end{align}
The rests are guaranteed by the Bianchi identity.
Explicit forms of the energy-momentum tensor and the charge densities are given in Appendix.
Then, we have
\begin{align}
  \frac{2m'}{r^2}-8\pi G\eta^2\biggl(\frac{e^2f^2\alpha^2}{\sigma^2(1-2m/r)}
	&+\frac{(e\alpha-\Omega)^2u^2}{\sigma^2(1-2m/r)}
	+\left(1-\frac{2m}{r}\right)( f'^2 +u'^2)
\cr
	&+\frac{\lambda}{4}(f^2-1)^2+\mu f^2u^2 +\frac{1}{2\sigma^2}\alpha'^2 \biggr)=0,
\label{eq:equation_Rtt}
\end{align}
and
\begin{align}
	\frac{(1-2m/r)\sigma'}{r\sigma}
		-8\pi G\eta^2\bigg( \frac{e^2f^2\alpha^2}{\sigma^2(1-2m/r)}
		&+\frac{(e\alpha-\Omega)^2u^2}{\sigma^2(1-2m/r)}
\cr
  	& +\left(1-\frac{2m}{r}\right) (f'^2 +u'^2)\bigg)=0.
\label{eq:equation_RttRrr}
\end{align}
The dimensionless parameter $G\eta^2=\left(\eta/M_\text{P}\right)^2$ represents
the symmetry breaking scale with respect to the Planck mass, $M_\text{P}$.
In the limiting case of $G\eta^2\to 0$, the gravitational field decouples
with the matter fields, where
the matter system \eqref{eq:equation_u}-\eqref{eq:equation_alpha}
with $\sigma(r)=1$ and $m(r)=0$
are studied in Refs.\cite{Ishihara:2018rxg,Ishihara:2019gim,Ishihara:2021iag}.

We require regularity of the spherically symmetric fields at the origin described by
\begin{align}
	\sigma'=0, \quad m=0, \quad u'=0, \quad f'=0, \quad \alpha'=0.
\label{eq:BC_origin}
\end{align}
In addition, we assume that the solutions are localized in finite regions.
Therefore, for the matter fields, we assume
\begin{align}
	u=0, \quad f=1, \quad \alpha=0,
\label{eq:BC_infinity_matter}
\end{align}
at spatial infinity.
The energy-momentum tensor $T_{\mu\nu}$ of the matter fields satisfying \eqref{eq:BC_origin}
and \eqref{eq:BC_infinity_matter} is localized in a neighborhood
of the origin with a quick radial decay,
then we require that the gravitational fields satisfy
\begin{align}
	\sigma=1, \quad m=m_{\infty}=\text{const}.
\label{eq:BC_infinity_metric}
\end{align}
at spatial infinity.

\section{Numerical Calculations}

In this section, we obtain solutions by solving the set of equations
\eqref{eq:equation_u}, \eqref{eq:equation_f}, \eqref{eq:equation_alpha},
\eqref{eq:equation_Rtt}, and \eqref{eq:equation_RttRrr}, numerically,
and study properties of the numerical soutions.
In this article, we fix the coupling constants as $e=1.0$, $\mu=1.4$, and $\lambda=1.0$,
for an example.
On the other hand, we consider various symmetry breaking scales.
For the Planck breaking scale, $\eta/M_\text{P}= 1$, we have
$m_{\psi}\sim m_{\phi}\sim m_A\sim 10^{19}\text{GeV}$, and for the lower breaking scale,
$\eta/M_\text{P} = 10^{-3}$, $m_{\psi}\sim m_{\phi}\sim m_A \sim 10^{16}\text{GeV}$, respectively.

At a large distance, $u(r)$ should decrease quickly
so that the energy is localized in a compact region.
The boundary conditions at the spatial infinity,
\eqref{eq:BC_infinity_matter} and \eqref{eq:BC_infinity_metric},
require $f=1, \alpha=0$, and $\sigma=1, m=\text{const}$,
then the equation \eqref{eq:equation_u} for $u(r)$ reduces to
\begin{align}
	u''-(\mu - \Omega^2 ) u = 0.
\end{align}
Then, $\Omega$ is bounded above by $\Omega_\text{max}:=\sqrt{\mu}$.
There is also the lower bound $\Omega_\text{min}$, which depends on $\eta/M_\text{P}$,
for existence of a solution, and, as is seen later,
the solution has the maximum mass for $\Omega$ near $\Omega_\text{min}$.
We can find numerical solutions for the parameter $\Omega$ in the range
\begin{align}
	\Omega_\text{min} \leq \Omega < \Omega_\text{max}.
\label{Omega_range}
\end{align}

Typical behaviors of the fields obtained by numerical calculations are shown
as functions of $r$ in Fig.\ref{fig:comfiguration_G1} for the Planck breaking scale case,
and in Fig.\ref{fig:comfiguration_G10em6} for the lower breaking scale case.
They show that the matter fields and the gravitational field are localized
in a finite region in each case.
Thus, they represent nontopological solitons with self-gravity,
we call them nontopological soliton star (NTS star) solutions.

\def\figwidth{5cm}

\begin{figure}[h]
\begin{tabular}{ccc}
\begin{minipage}{0.5\hsize}
\centering
\text{Planck breaking scale: $\eta/M_\text{P}=1$} \par\bigskip
\end{minipage}
\\
\begin{minipage}{0.33\hsize}
\centering
\text{(a) $\Omega=1.183$}\par\medskip
\includegraphics[width=\figwidth]{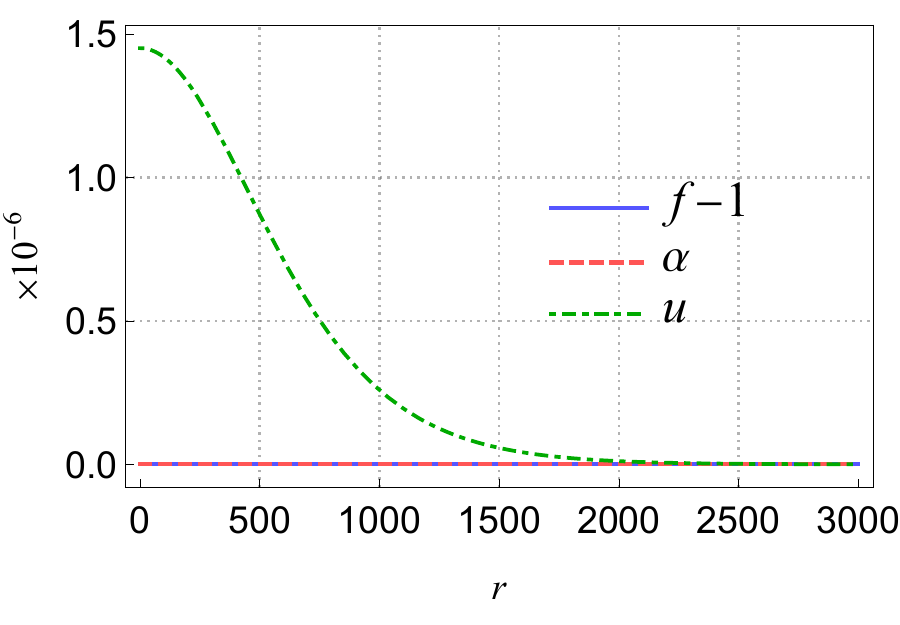}\\
\includegraphics[width=\figwidth]{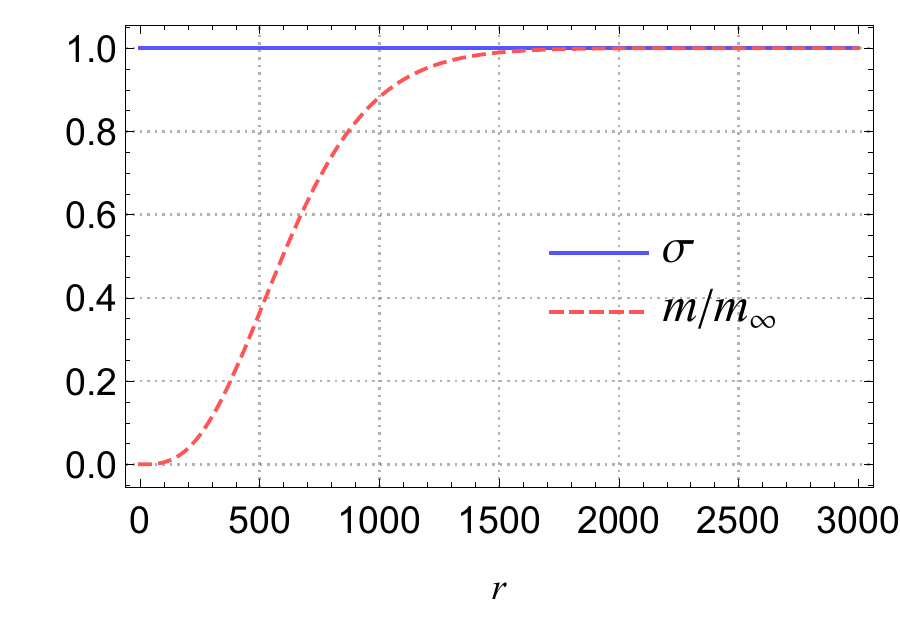}
\end{minipage}
\begin{minipage}{0.33\hsize}
\centering
\text{(b) $\Omega=1.178$}\par\medskip
\includegraphics[width=\figwidth]{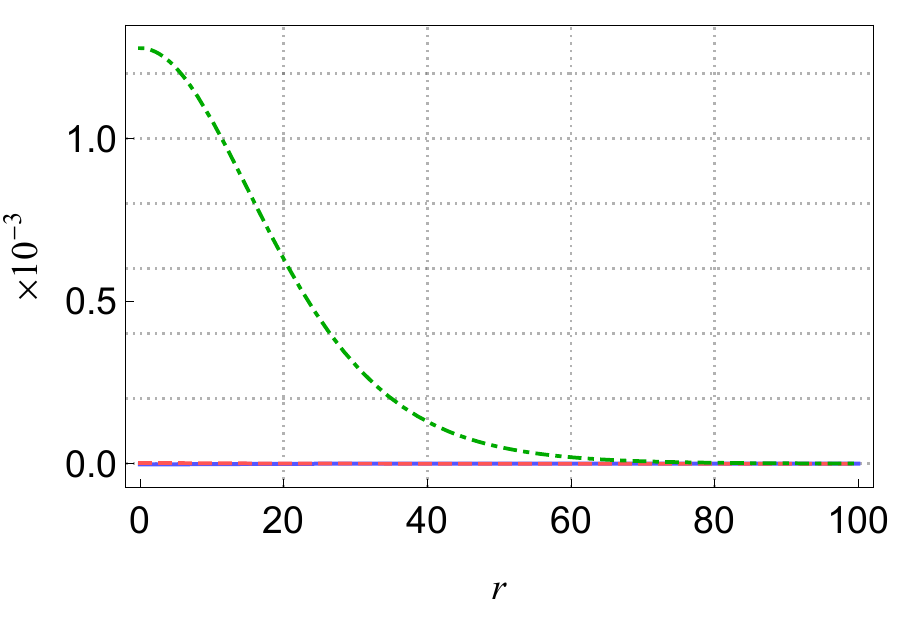}\\
\includegraphics[width=\figwidth]{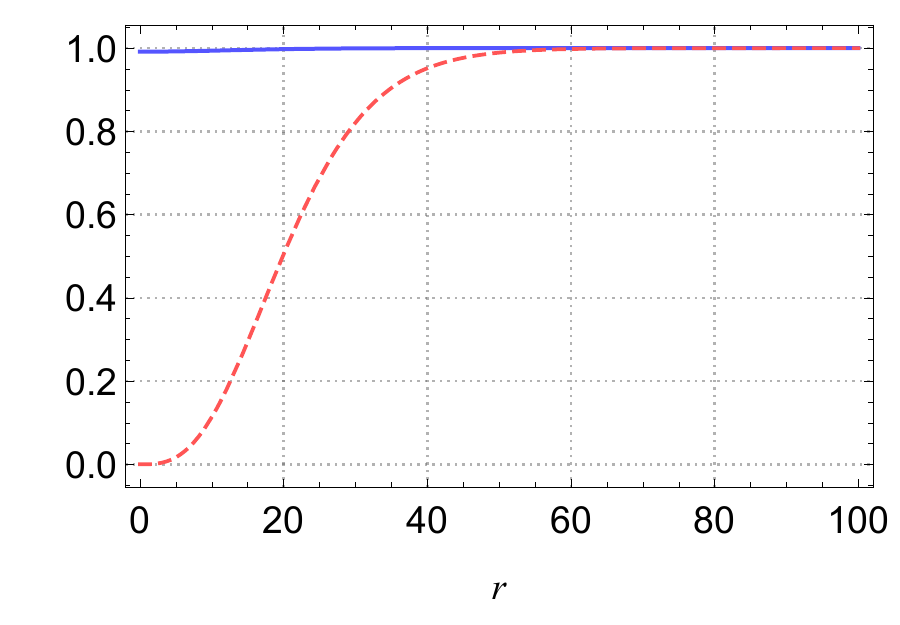}
\end{minipage}
\begin{minipage}{0.33\hsize}
\centering
\text{(c) $\Omega=1.008$}\par\medskip
\includegraphics[width=\figwidth]{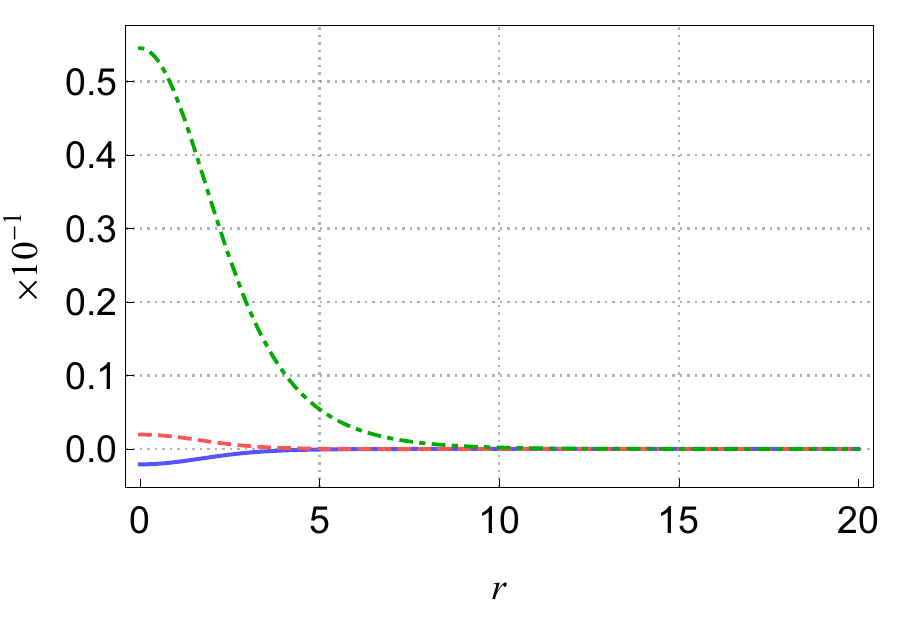}\\
\includegraphics[width=\figwidth]{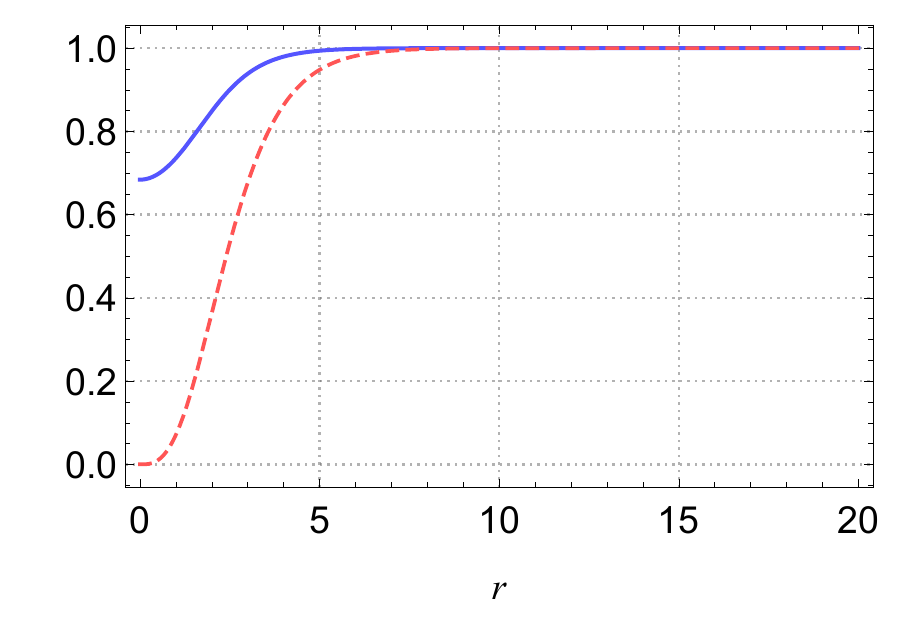}
\end{minipage}
\\
\end{tabular}
\caption{
Field configurations of numerical solutions in the Planck breaking scale case
$\eta/M_\text{P}=1$.
We show three subcases of the parameter:
(a) $\Omega=1.183$ (left column), (b) $\Omega=1.178$ (middle column),
and (c) $\Omega=1.008$ (right column).
The scalar fields $u, f$ and the gauge field $\alpha$ are plotted in the upper pannels,
and the metric components $\sigma$ and $m$ are plotted in the lower panels.
\label{fig:comfiguration_G1}
}
\end{figure}

\begin{figure}[H]
\begin{tabular}{ccc}
\begin{minipage}{0.5\hsize}
\centering
\text{Lower breaking scale: $\eta/M_\text{P}=10^{-3}$}\par\bigskip
\end{minipage}
\\
\begin{minipage}{0.33\hsize}
\centering
\text{(d) $\Omega=1.183$}\par\medskip
\includegraphics[width=\figwidth]{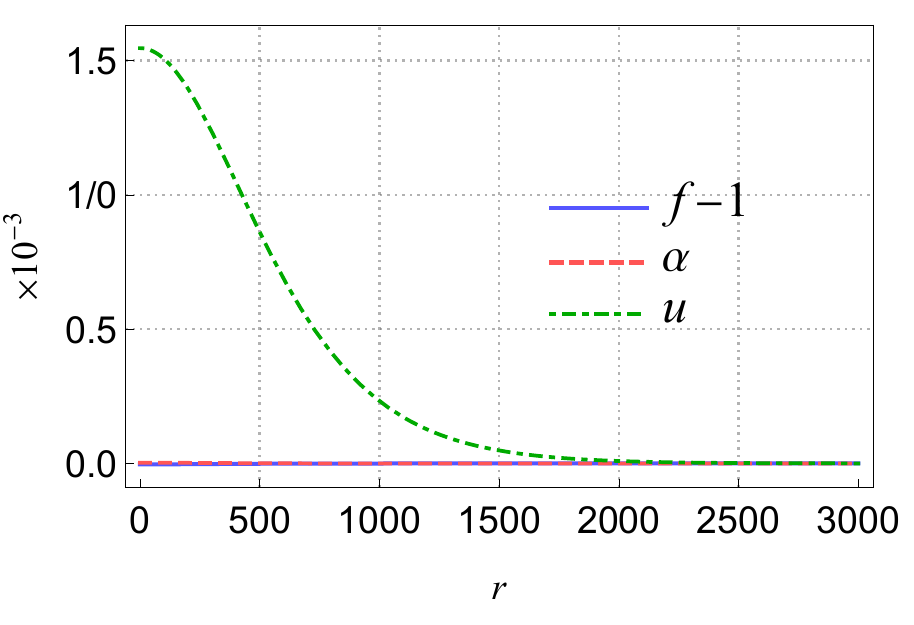}\\
\includegraphics[width=\figwidth]{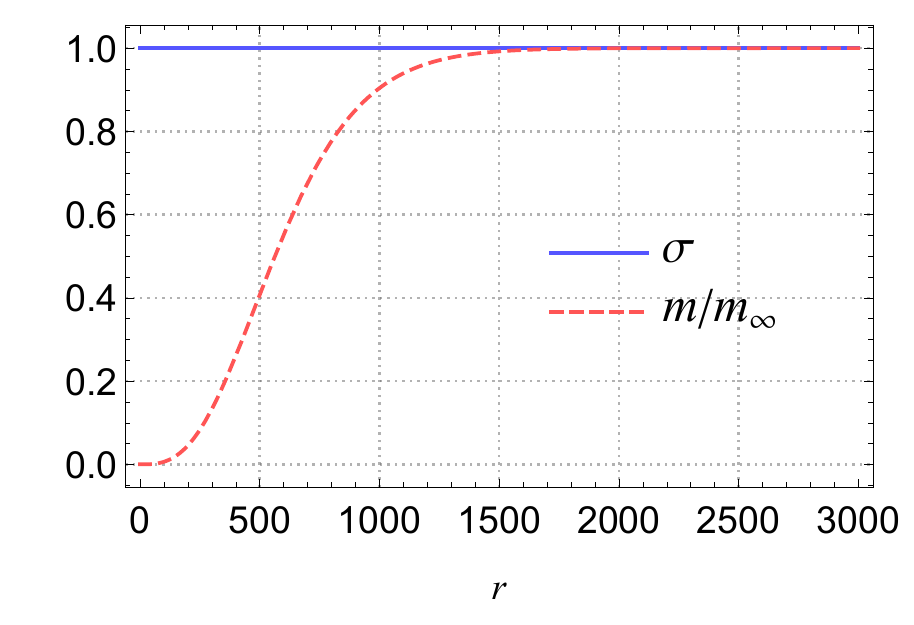}
\end{minipage}
\begin{minipage}{0.33\hsize}
\centering
\text{(e) $\Omega=1.178$}\par\medskip
\includegraphics[width=\figwidth]{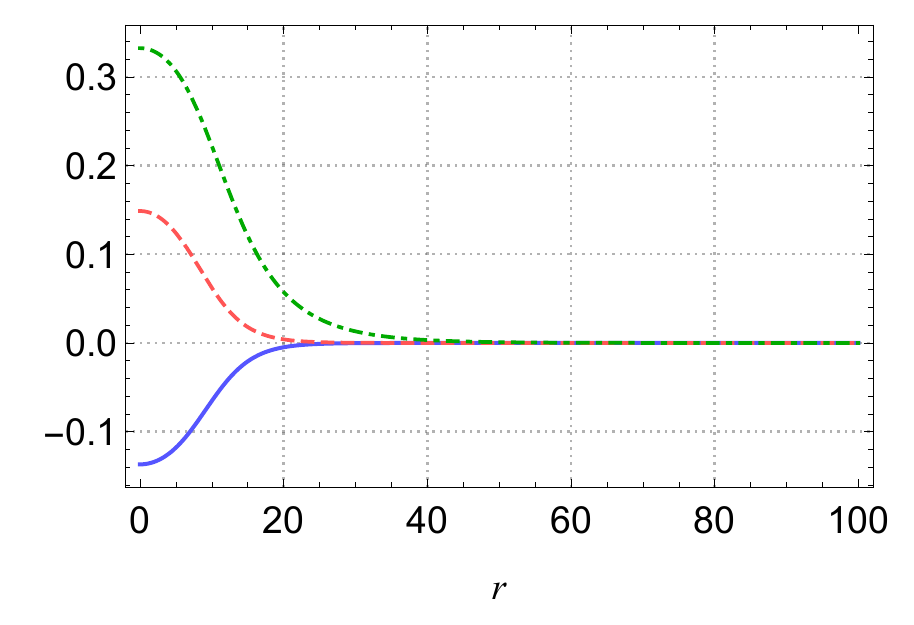}\\
\includegraphics[width=\figwidth]{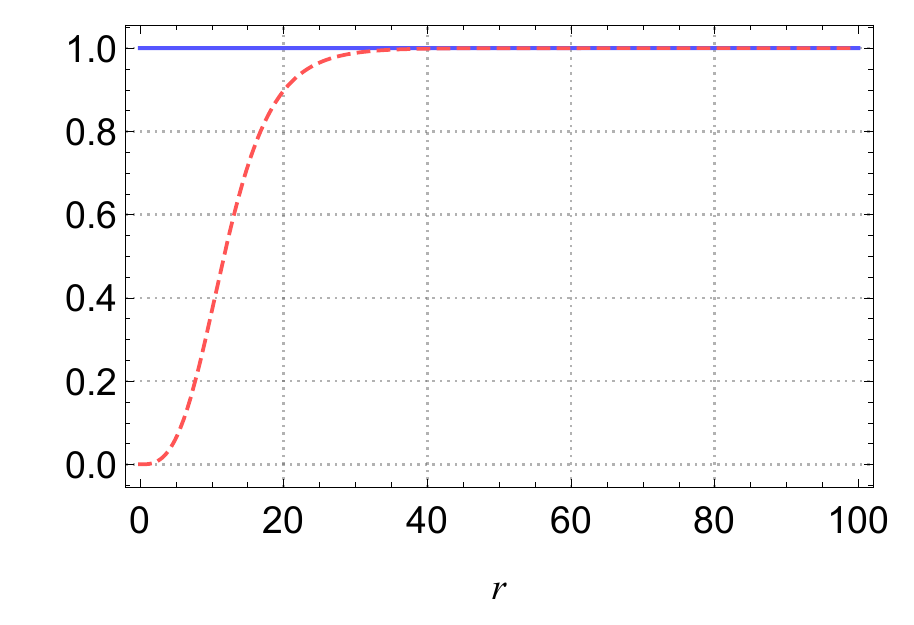}
\end{minipage}
\begin{minipage}{0.33\hsize}
\centering
\text{(f) $\Omega=0.783$}\par\medskip
\includegraphics[width=\figwidth]{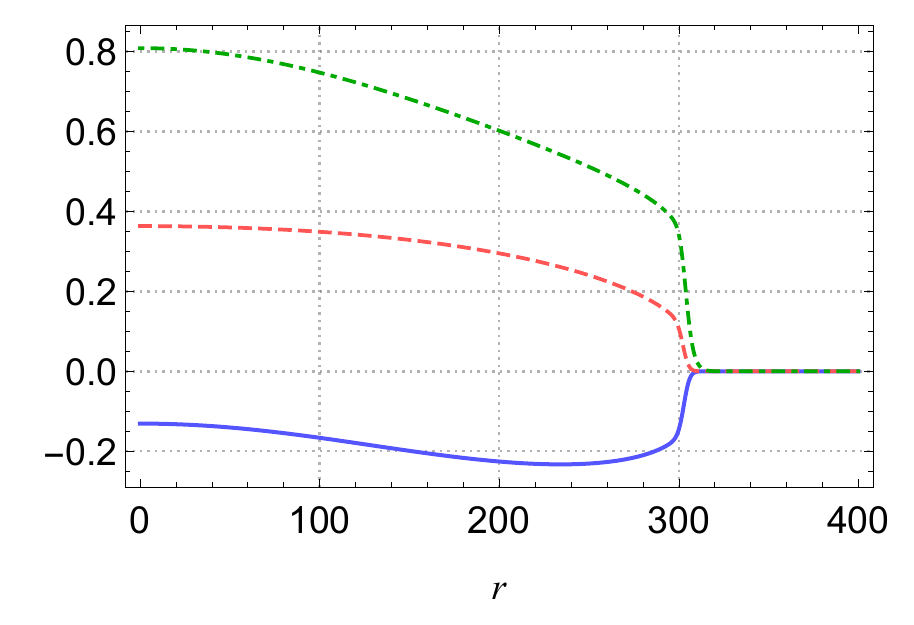}\\
\includegraphics[width=\figwidth]{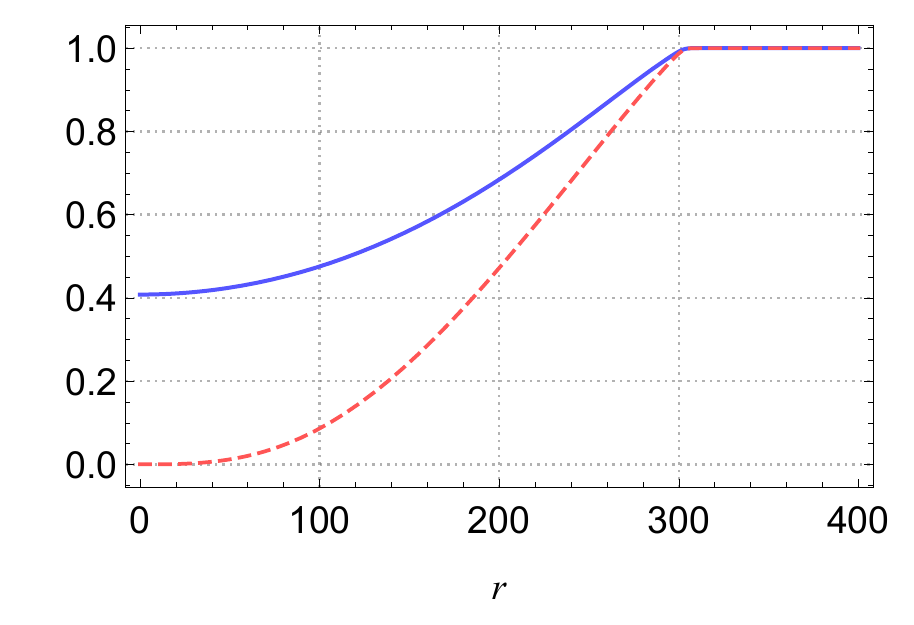}
\end{minipage}
\end{tabular}
\caption{
The same ones as Fig.\ref{fig:comfiguration_G1} in the lower breaking scale,
$\eta/M_\text{P}=10^{-3}$.
We show for (d) $\Omega=1.183$ (left column), (e) $\Omega=1.178$ (middle column),
and (f) $\Omega=0.783$  (right column).
\label{fig:comfiguration_G10em6}
}
\end{figure}

In the cases (a) and (b)  in the Planck breaking scale, and (d) in the lower breaking scale,
the function $u$
has Gaussian forms, while $f\sim 1$ and $\alpha\sim 0$ almost everywhere.
In these cases, the scalar field $\phi$ and the gauge field $A$ are not excited anywhere,
and only the scalar field $\psi$, which has the mass $m=m_{\psi}$ by the Higgs mechanism,
becomes a source of gravity.
The behavior that the massive scalar field with gravity yields compact objects
is just the same as \lq mini boson star\rq\ solutions \cite{Kaup:1968zz,Ruffini:1969qy}.
On the other hand, in the cases (e) and (f) in the lower breaking scale,
$\psi, \phi$, and $A$ are exited inside the stars.
In these cases, interaction between the scalar fields and the gauge field plays
an important role for appearance of solutions \cite{Ishihara:2018rxg,Ishihara:2019gim,Ishihara:2021iag}.

As for the gravity,
in the cases (a), (b), (d), and (e), the lapse function, $\sigma$, is almost constant.
It means that the gravity is weak so that the the Newtonian description is possible.
In contrast, in the cases (c) and (f), $\sigma$ varies significantly from $r=0$ to infinity,
i.e., it means the gravity requires relativistic description.
Paying attention to the scale of horizontal axises, we see the size of the NTS stars depend
on $\Omega$. The dependence is discussed later in detail.

\section{Energy density, Pressure, and Charge densities}

\subsection{Energy density and pressure }

The energy density and the pressure, defined by
\eqref{eq:dimensionless_Ttt}-\eqref{eq:dimensionless_Tthetatheta} in Appendix \ref{appendix},
are plotted in Fig.\ref{fig:MassPressure} for numerical solutions.
We see that the pressures can be ignored compared to the energy density
in the cases (a), (b) in the Planck breaking scale case, and
in the cases (d), (e) in the lower breaking scale case.
Then, NTS stars in these cases behave as \lq gravitating dust balls\rq .
On the other hand, in the cases (c) and (f),
the radial and tangential pressures become large in the central regions.

\subsection{Charge distribution}

We plot the charge density of $\psi$, $\rho_{\psi}$, and charge density of $\phi$, $\rho_{\phi}$, and total one $\rho_\text{total}=\rho_{\psi}+\rho_{\phi}$
of the NTS stars in Fig.\ref{fig:charge}.
In all cases except (c), $\rho_{\psi}$ is compensated
by $\rho_{\phi}$ then charge screening occurs everywhere \cite{Ishihara:2018rxg}.
In the case (c), $\rho_\text{total}$ is positive in the central region,
and negative surround the region. Total charge, integration of $\rho_\text{total}$ from $r=0$
to the large $r$, vanishes. Namely, the charge is totally screened. This fact is consistent
with that the gauge field, $A$, becomes massive, and $\alpha$ decays quickly as $r\to\infty$.

\def\figwidth{4.4cm}

\begin{figure}[H]
\begin{tabular}{ccc}
\begin{minipage}{0.5\hsize}
\centering
\text{Planck breaking scale: $\eta/M_\text{P}=1$} \par\bigskip
\end{minipage}
\\
\begin{minipage}{0.33\hsize}
\centering
\text{(a) $\Omega=1.183$}\par\medskip
\includegraphics[width=\figwidth]{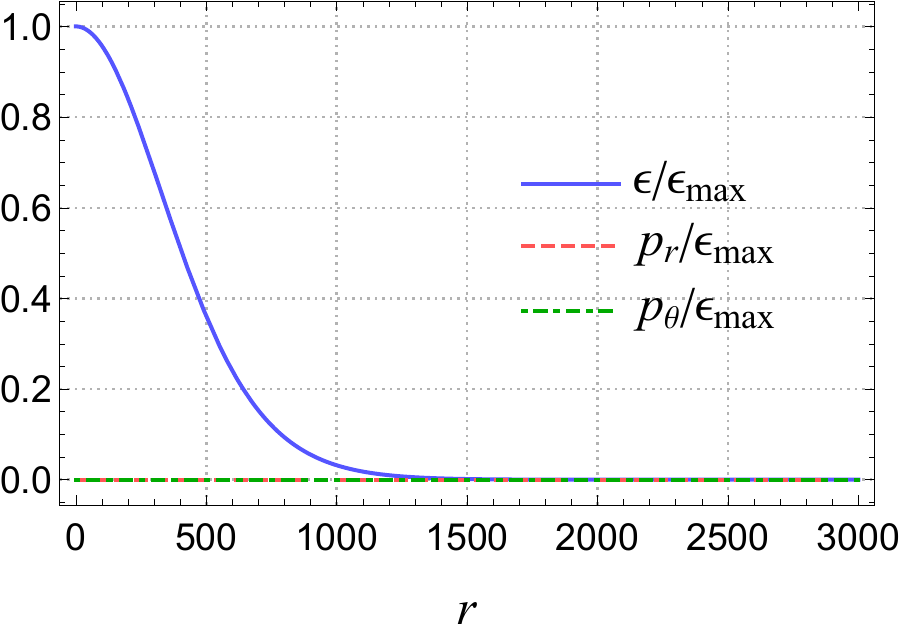}
\end{minipage}
\begin{minipage}{0.33\hsize}
\centering
\text{(b) $\Omega=1.178$}\par\medskip
\includegraphics[width=\figwidth]{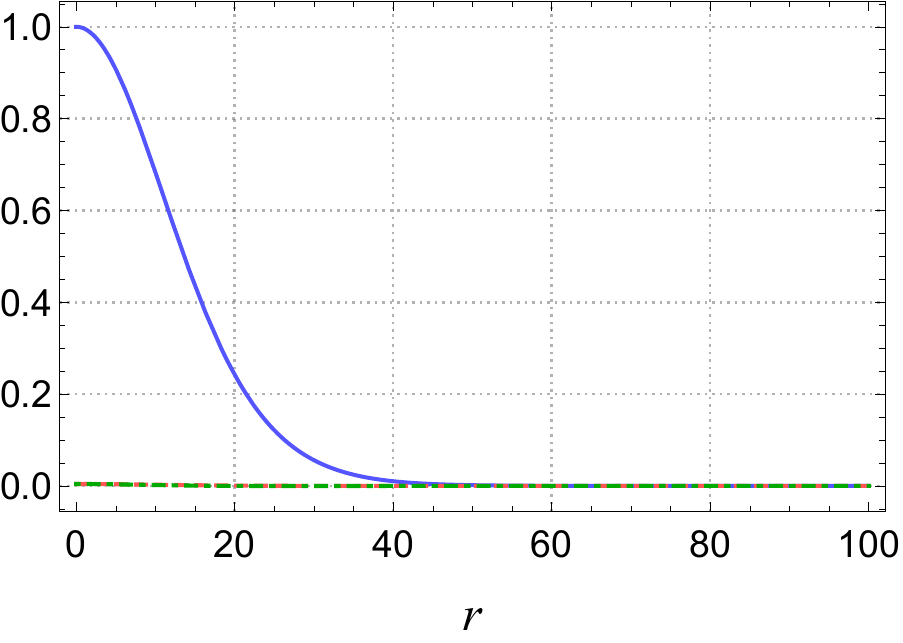}
\end{minipage}

\begin{minipage}{0.33\hsize}
\centering
\text{(c) $\Omega=1.008$}\par\medskip
\includegraphics[width=\figwidth]{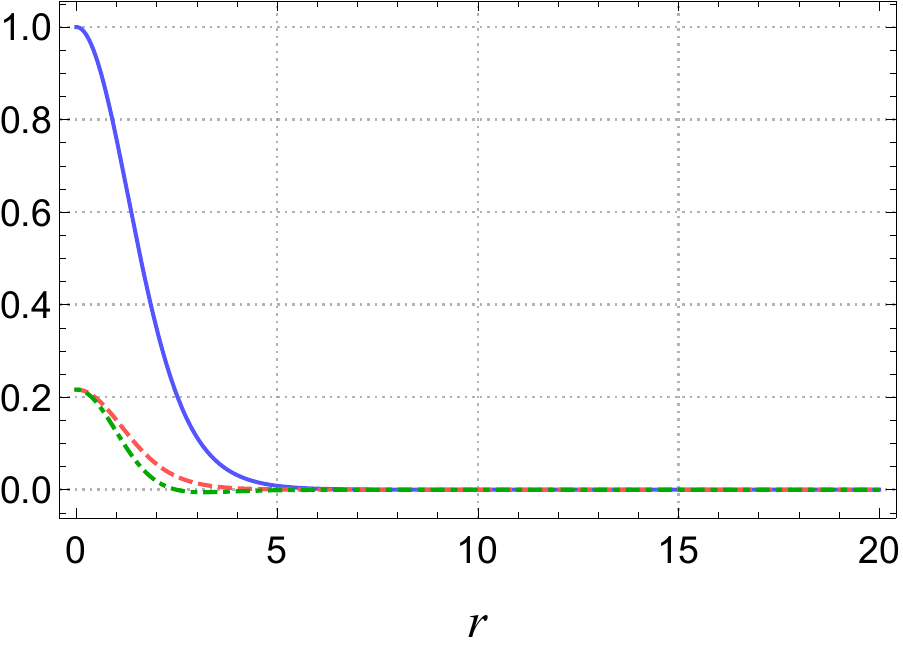}
\end{minipage}
\vspace{5mm}
\\
\begin{minipage}{0.5\hsize}
\centering
\text{Lower breaking scale: $\eta/M_\text{P}=10^{-3}$}\par\bigskip
\end{minipage}
\\
\begin{minipage}{0.33\hsize}
\centering
\text{(d) $\Omega=1.183$}\par\medskip
\includegraphics[width=\figwidth]{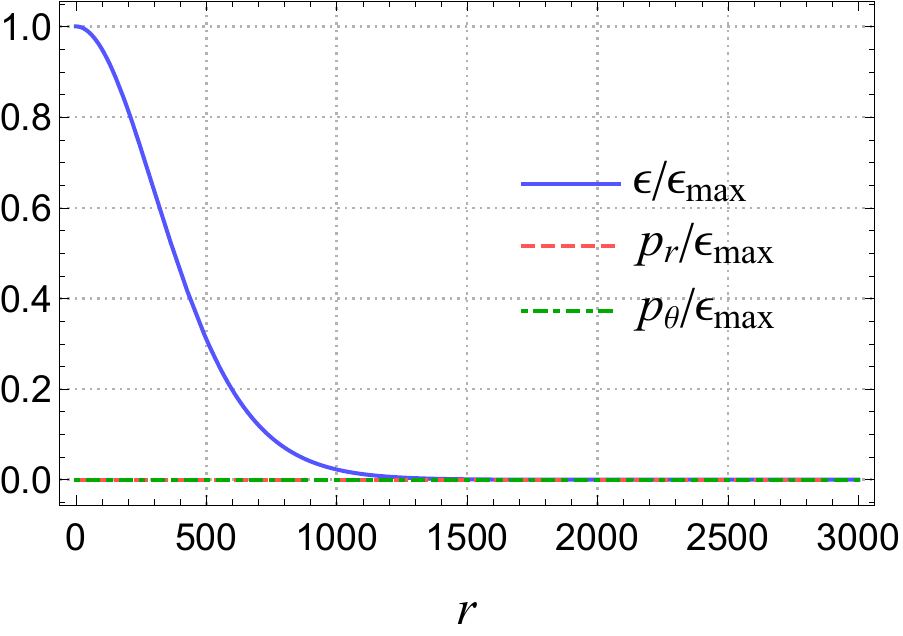}
\end{minipage}
\begin{minipage}{0.33\hsize}
\centering
\text{(e) $\Omega=1.178$}\par\medskip
\includegraphics[width=\figwidth]{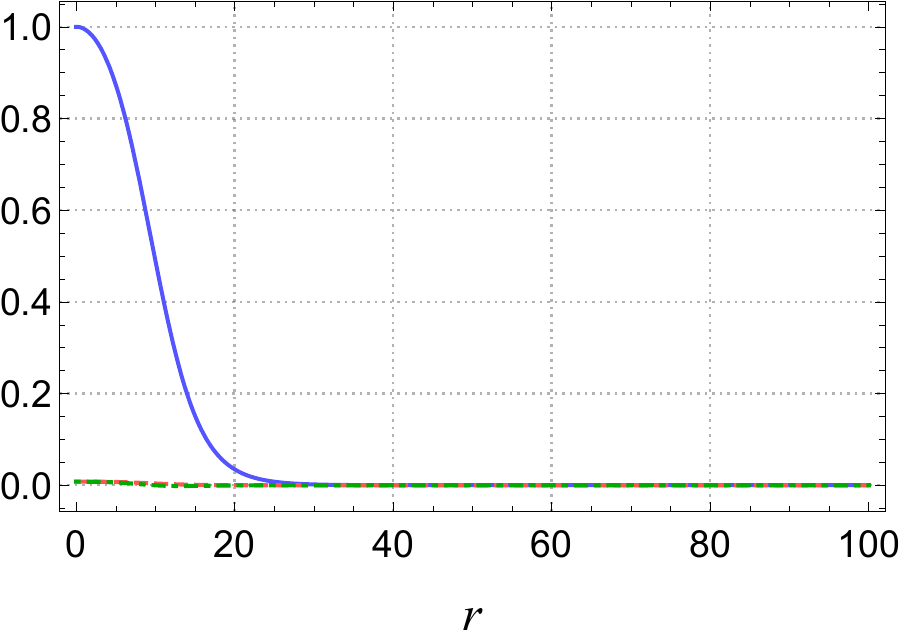}
\end{minipage}
\begin{minipage}{0.33\hsize}
\centering
\text{(f) $\Omega=0.783$}\par\medskip
\includegraphics[width=\figwidth]{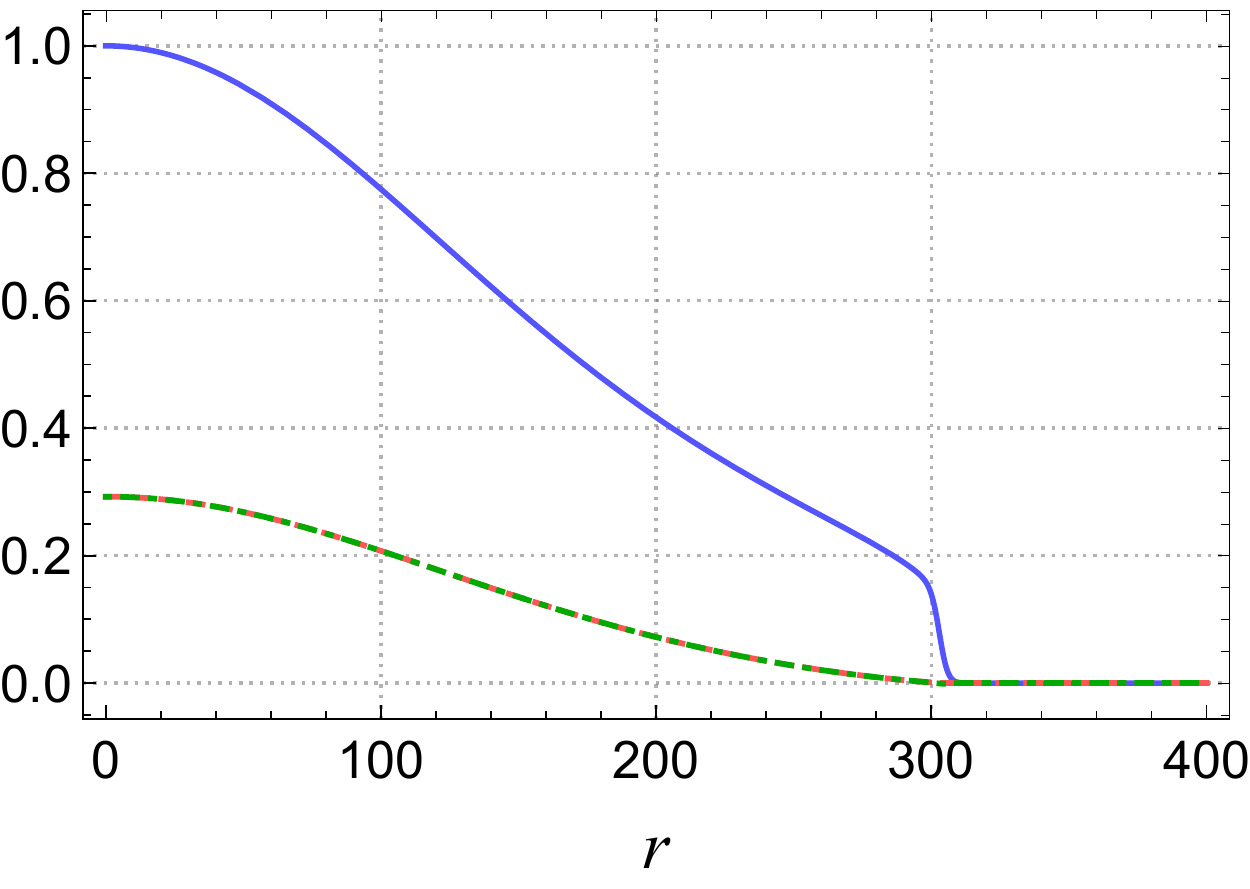}
\end{minipage}
\end{tabular}
\caption{
The energy density $\epsilon$,
the radial pressure $p_r$,
and the tangential pressure $p_{\theta}$,
where $\epsilon_\text{max}$ is the maximum value of $\epsilon$.
The upper panels correspond to the Planck breaking scale case, $\eta/M_\text{P}=1$,
and the lower does the lower breaking scale case, $\eta/M_\text{P}=10^{-3}$.
}
\label{fig:MassPressure}
\end{figure}

\begin{figure}[H]
\begin{tabular}{ccc}
\begin{minipage}{0.5\hsize}
\centering
\text{Planck breaking scale: $\eta/M_\text{P}=1$}\par\bigskip
\end{minipage}
\\
\begin{minipage}{0.33\hsize}
\centering
\text{(a) $\Omega=1.183$}\par\medskip
\includegraphics[width=\figwidth]{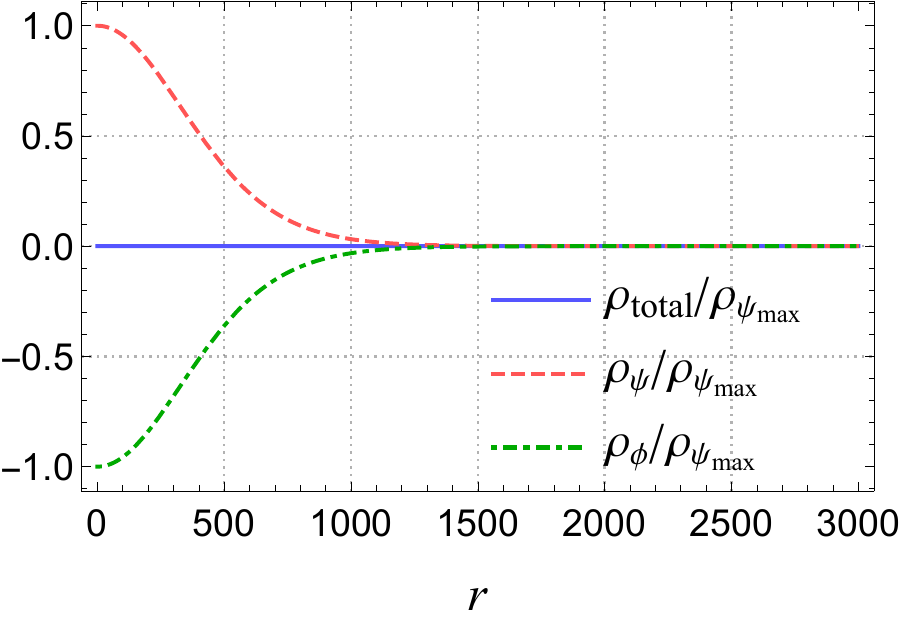}
\end{minipage}
\begin{minipage}{0.33\hsize}
\centering
\text{(b) $\Omega=1.178$}\par\medskip
\includegraphics[width=\figwidth]{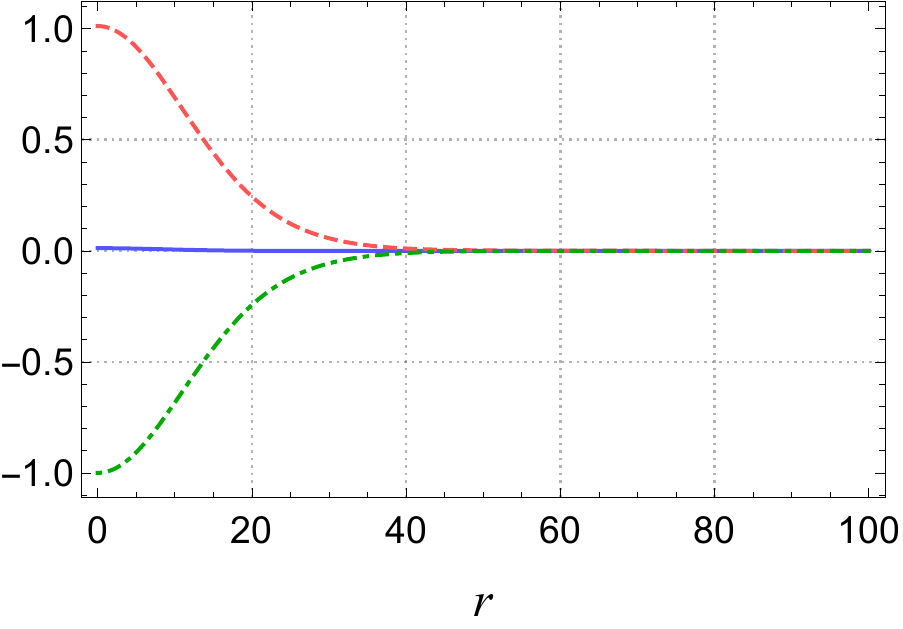}
\end{minipage}
\begin{minipage}{0.33\hsize}
\centering
\text{(c) $\Omega=1.008$}\par\medskip
\includegraphics[width=\figwidth]{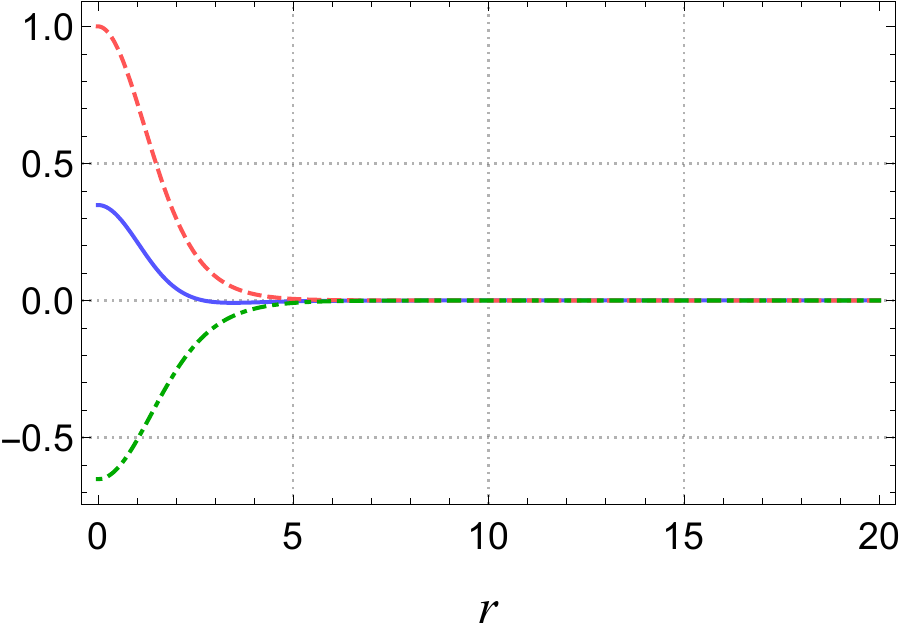}
\end{minipage}
\vspace{5mm}
\\
\begin{minipage}{0.5\hsize}
\centering
\text{Lower breaking scale: $\eta/M_\text{P}=10^{-3}$}\par\bigskip
\end{minipage}
\\
\begin{minipage}{0.33\hsize}
\centering
\text{(d) $\Omega=1.183$}\par\medskip
\includegraphics[width=\figwidth]{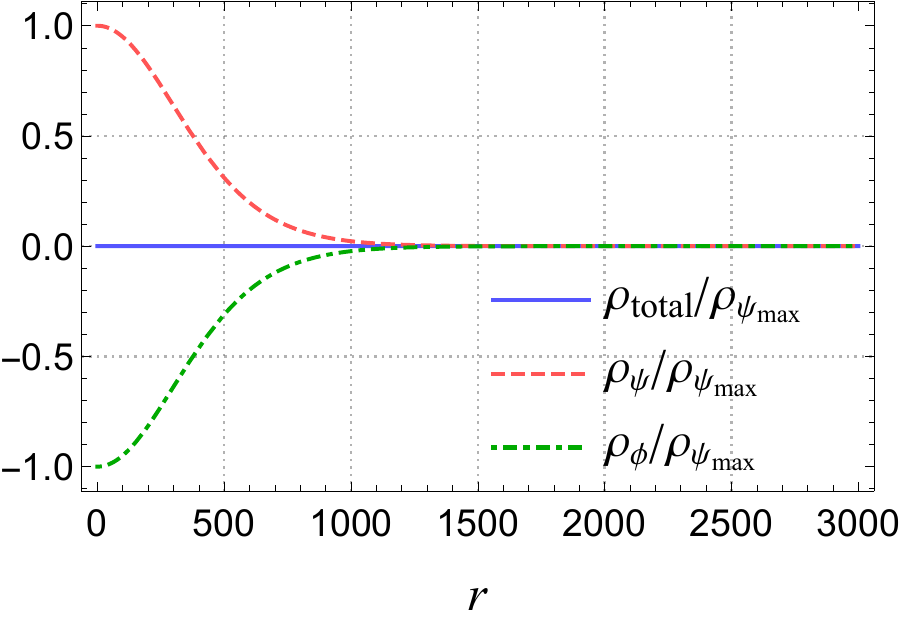}
\end{minipage}
\begin{minipage}{0.33\hsize}
\centering
\text{(e) $\Omega=1.178$}\par\medskip
\includegraphics[width=\figwidth]{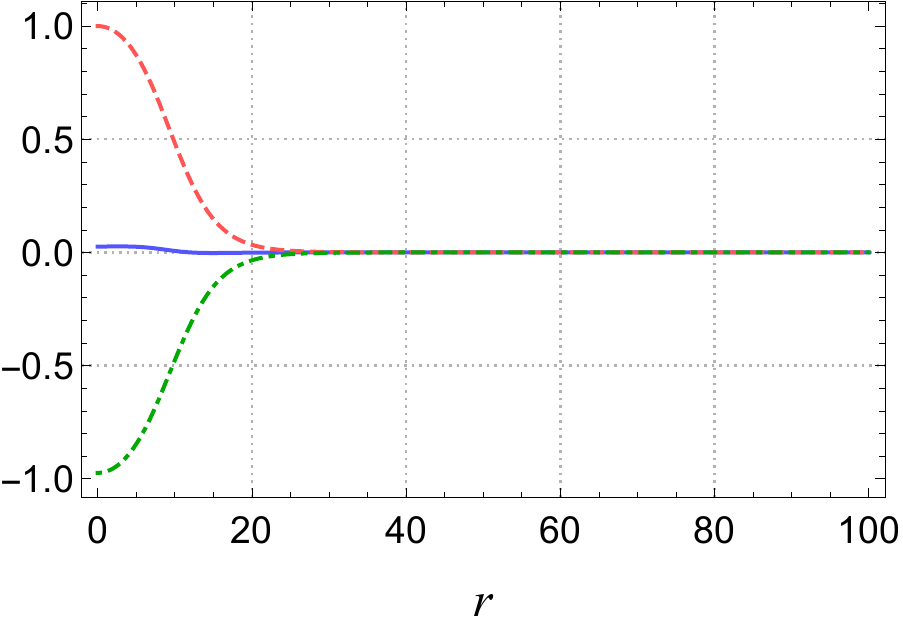}
\end{minipage}
\begin{minipage}{0.33\hsize}
\centering
\text{(f) $\Omega=0.783$}\par\medskip
\includegraphics[width=\figwidth]{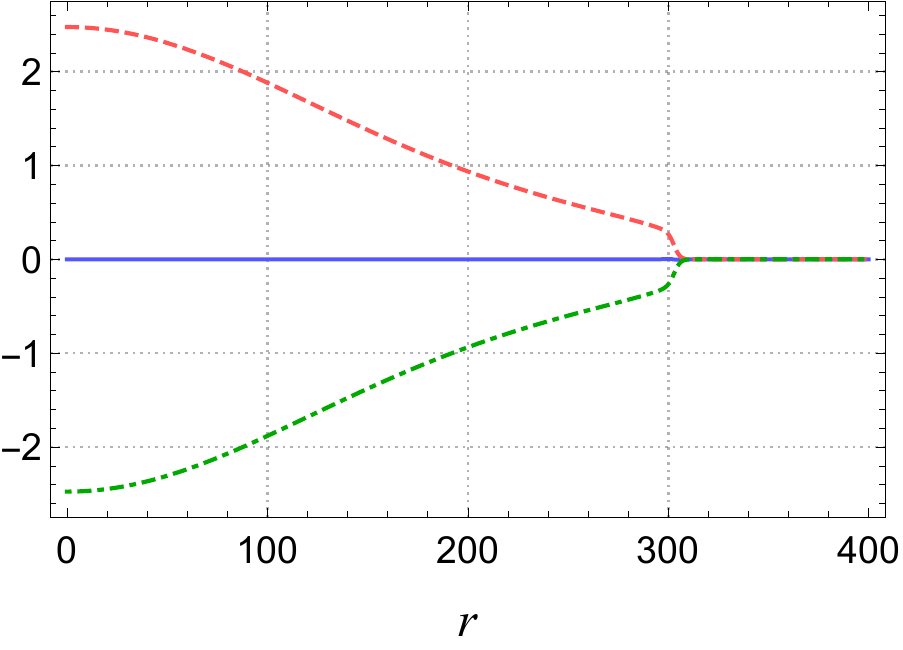}
\end{minipage}
\end{tabular}
\caption{
The charge densities $\rho_{\psi}$, $\rho_{\phi}$,
and the total charge density $\rho_\text{total}=\rho_{\psi}+\rho_{\phi}$ are plotted,
where these charge densities are normalized by the maximum value of $\rho_{\psi}$,
$\rho_{\psi_\text{max}}$.
The upper panels correspond to the Planck breaking scale case, $\eta/M_\text{P}=1$,
and the lower does the lower breaking scale case, $\eta/M_\text{P}=10^{-3}$.
}
\label{fig:charge}
\end{figure}

\section{Mass and radius}

We see in Figs.\ref{fig:comfiguration_G1}, \ref{fig:comfiguration_G10em6}, \ref{fig:MassPressure},
that the size of NTS solutions much depend on the parameter $\Omega$.
We study total mass and radius of the solutions in this section.

\subsection{Gravitational Mass}

\begin{figure}[H]
\centering
\includegraphics[width=10cm]{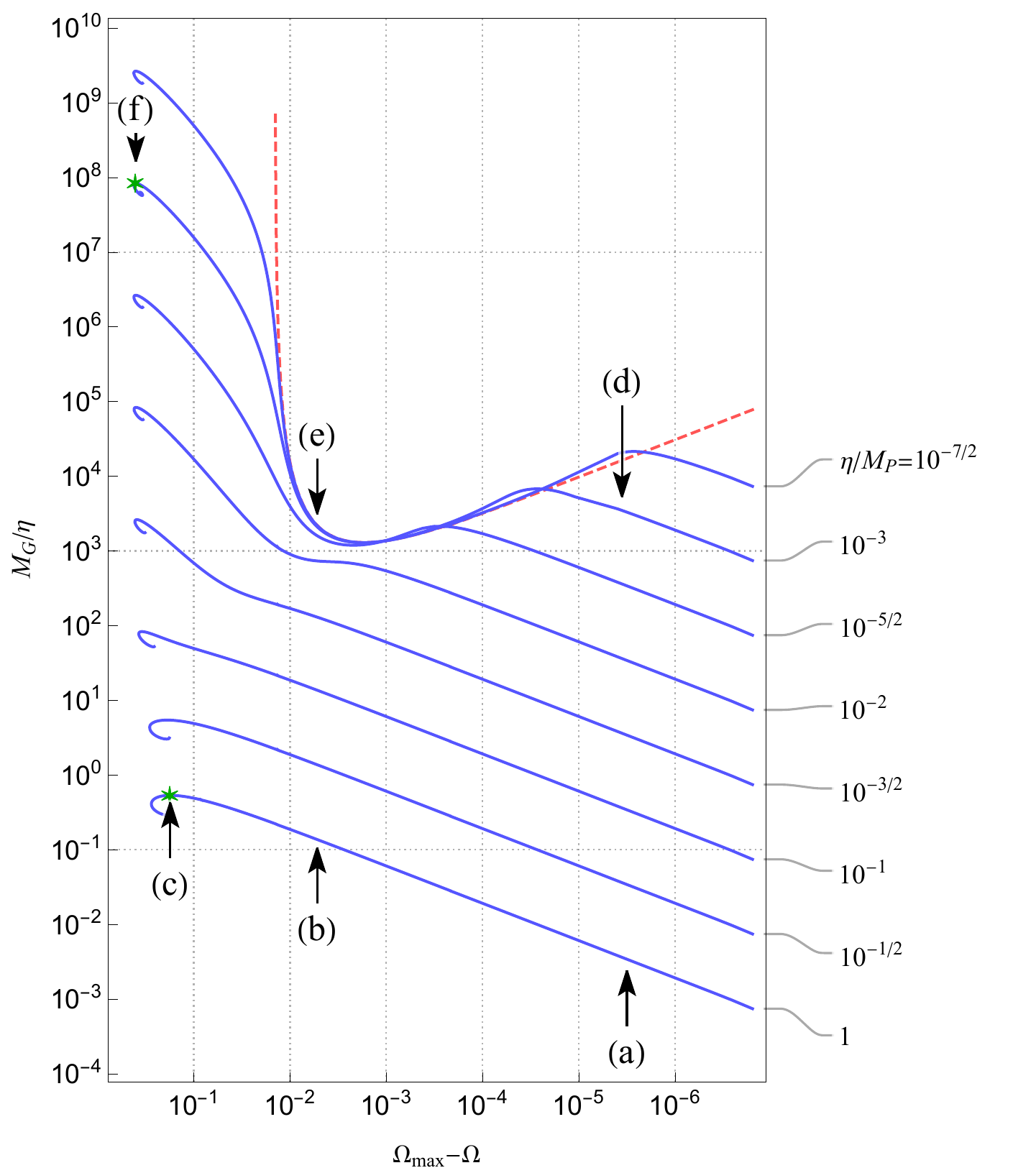}
  \hfill
\caption{
The gravitational mass of NTS stars as a function of $\Omega$
for various breaking scales $\eta/M_\text{P}$.
The vertical axis is taken for $M_G/\eta$
and the horizontal axis for $\log{(\Omega_\text{max}-\Omega)}$.
The (red) broken curve denotes the mass of NTS solutions, dust balls \cite{Ishihara:2019gim,Ishihara:2021iag}, decouple to gravity,
i.e., $\eta/M_\text{P}\to 0$. The points (a) - (f) correspond to the solutions shown in
Figs.\ref{fig:comfiguration_G1}-\ref{fig:charge}
}
\label{fig:Massprofile}
\end{figure}

The gravitational mass
of the NTS star, $M_G$, is given by
\begin{align}
	 M_G
	= \frac{m_\infty}{G\eta}.
\end{align}
For a fixed $\eta/M_\text{P}$, we have a NTS star solution for each $\Omega$,
then $M_G$ is a function of $\Omega$.
In Fig.\ref{fig:Massprofile}, curves represent $M_G$ as functions of $\Omega$
for various breaking scales $\eta/M_\text{P}$.
We find each curve has a spiral shape at the left end.
Then there appears lower limit of $\Omega$, $\Omega_\text{min}$,
for existence of  NTS star solutions.
Numerically, we have $\Omega_\text{min}\sim 0.91$ for $\eta/M_\text{P} \gtrsim 10^{-1}$,
while $\Omega_\text{min}\sim 0.765$ for $\eta/M_\text{P} \lesssim 10^{-2}$.
The gravitational mass $M_G$ is
multi-valued in $\Omega$ near a region $\Omega \sim \Omega_\text{min}$.
For a fixed $\eta/M_\text{P}$, there exists maximum of $M_G$ near $\Omega \sim \Omega_\text{min}$.
We call the NTS star with the maximum mass \lq{\sl maximum NTS star}\rq.
The maximum NTS stars for $\eta/M_\text{P}=10^{-3}$ and $\eta/M_\text{P}=1$ are marked by asterisks
in Fig.\ref{fig:Massprofile}.
In Fig.\ref{fig:MassPressure}, the central pressure in the maximum NTS star cases,
(c) and (f), become large in the order of $1/4\sim 1/3$ times the central energy density.
The pressure gradient balances to the gravitational force by the large mass.

\begin{figure}[H]
\centering
\includegraphics[width=10cm]{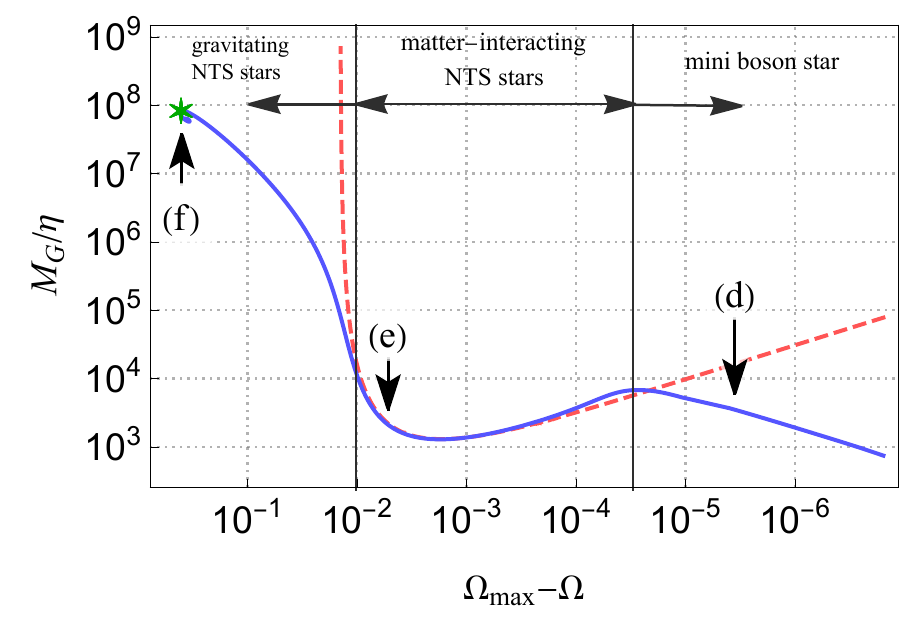}
\caption{
The gravitational mass of NTS stars
for the lower breaking scales $\eta/M_\text{P}=10^{-3}$.
In this case, the NTS star solutions are classified into three types:
mini boson stars, matter-interacting NTS stars, and gravitating NTS stars.
}
\label{fig:Massratio_loglog}
\end{figure}

For $\eta/M_\text{P} \gtrsim 10^{-1}$, the curves are shifted upward, as a whole,
as $\eta/M_\text{P}$ decreases,
while for $\eta/M_\text{P} \lesssim 10^{-2}$, the curves are modified,
and middle part of the curves converge to the limiting curve of $\eta/M_\text{P}=0$.
Let us pay attention to the curve of $\eta/M_\text{P} =10^{-3}$, for example (see Fig.\ref{fig:Massratio_loglog}).
The curve is divided into three parts: the middle segment that lies on the limiting
curve $\eta/M_\text{P} =0$,
the right segment apart downward from the limiting curve,
the left segment apart leftward.
Firstly, solutions on the right segment are \lq{\sl mini boson stars}\rq\ as mentioned before.
Secondly, solutions on the middle are NTS stars whose gravity can be neglected,
while the interaction of the scalar fields and the gauge field is important,
then we call them \lq{\sl matter-interacting NTS stars}\rq.
Thirdly, solutions on the left segment are NTS stars whose gravity is important,
then we call them \lq{\sl gravitating NTS stars}\rq.
In the family of the gravitating NTS stars, the mass quickly increases as $\Omega$ approaches
$\Omega_\text{min}$.
The points (a) - (f)  on the curves in Fig.\ref{fig:Massprofile} correspond
to the solutions shown in Figs.\ref{fig:comfiguration_G1} and \ref{fig:comfiguration_G10em6},
respectively.

\subsection{Surface Radius}

\begin{figure}[H]
\centering
\includegraphics[width=10cm]{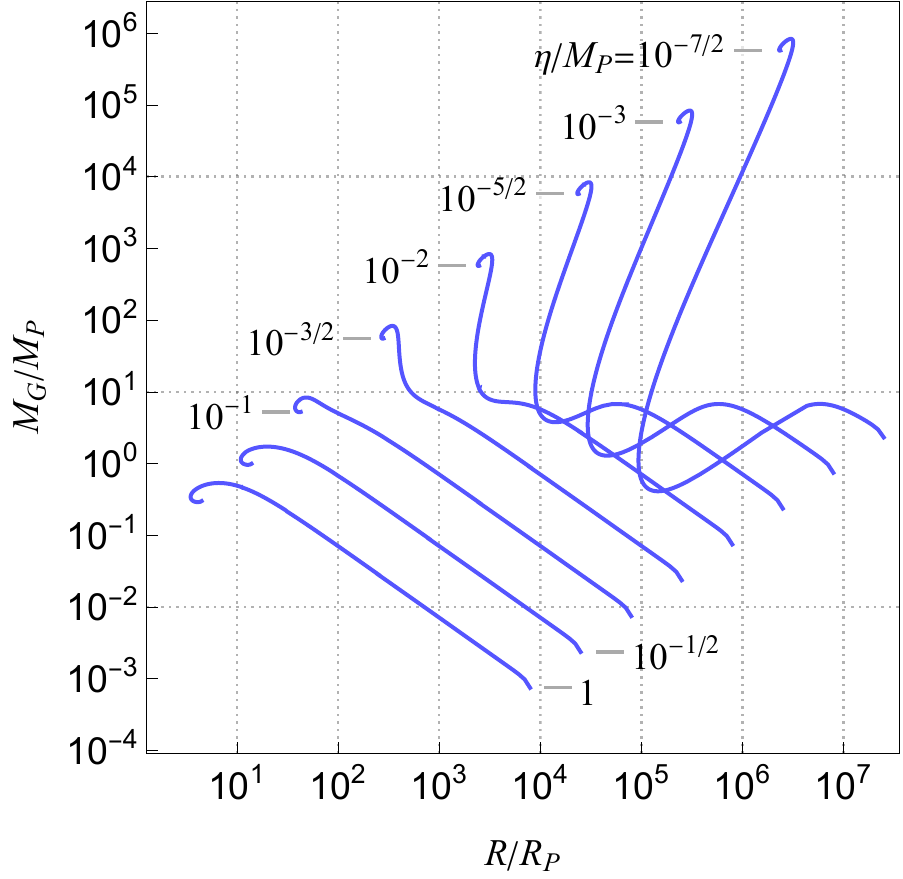}
\caption{
The gravitational mass $M_G$ of the NTS stars as a function of surface radius $R$
for various breaking scales.
In the vertical axis $M_G$ is normalized by the Planck mass $M_\text{P}$,
and in the horizontal axis $R$ is normalized by the Planck length $R_\text{P}:=\sqrt{G}$.
}
\label{fig:R99}
\end{figure}

We define the surface radius of the numerical solutions, say $r_s$, by
\begin{align}
	m(r_s):=0.99~ m_{\infty}.
\label{eq:numerical_R}
\end{align}
Namely, $99\%$ of total mass of the NTS stars includes within the surface radius $r_s$.
We plot gravitational mass $M_G$ of the NTS stars normalized by $M_\text{P}$ as a function of
dimensionful radius $R:=r_s/\eta$
for various breaking scales in Fig.\ref{fig:R99}.
For a fixed breaking scale in the range $\eta/M_\text{P} \gtrsim 10^{-1}$,
$M_G$ increases toward the maximum mass
as $R$ decreases,
while in the range $\eta/M_\text{P} \lesssim 10^{-5/2}$, $M_G$ depends on $R$ in a complicated way:
in the region $M_G/M_\text{P} \lesssim 10$, local maximum and local minimum appear and
in the region $M_G/M_\text{P} \gtrsim 10$,
$M_G$ increases toward the maximum mass as $R$ increases.

\subsection{Compactness}

Next, we investigate the compactness,
$C$, defined by
\begin{align}
 	C:=\frac{2 G M_G }{R}
	=\frac{2 m_{\infty}}{r_s} .
  \label{eq:compactness}
\end{align}
In the Schwarzschild geometries, which are exterior of NTS stars,
if $C\ge 2/3$ there exists the photon sphere and if $C\ge 1/3$ the innermost stable circular orbit
(ISCO) appears.

\begin{figure}[H]
\centering
\includegraphics[width=10cm]{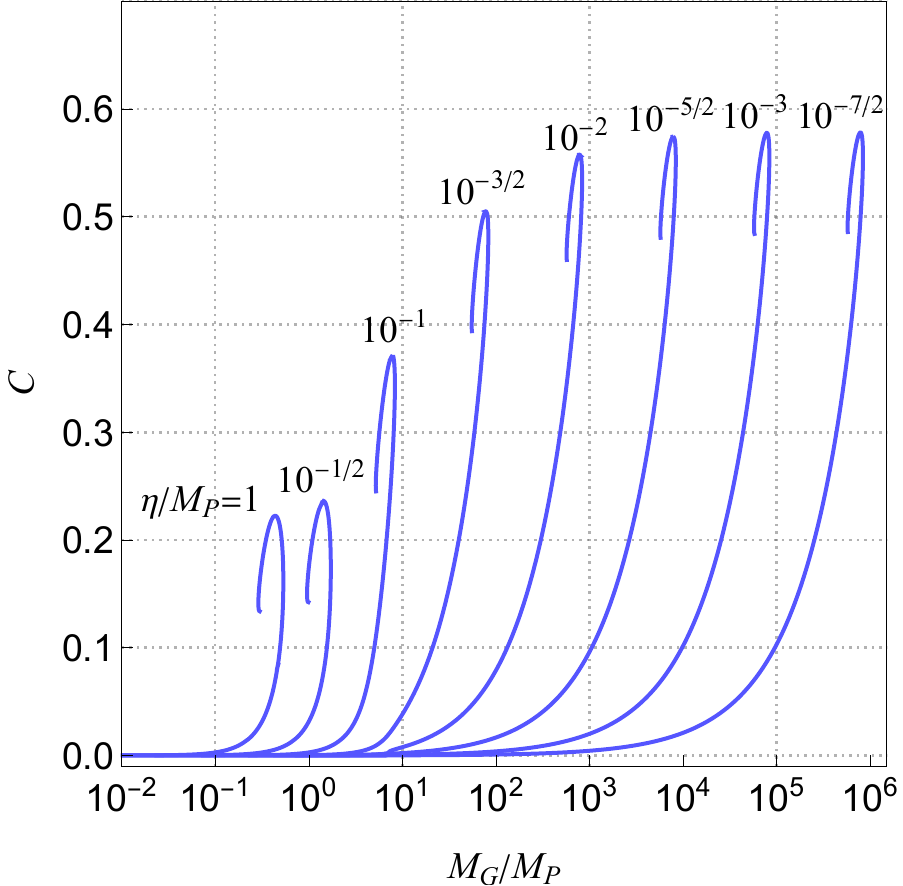}
\caption{
The compactness of NTS stars as a function of the gravitational mass $M_G$
for various breaking scales.
}
\label{fig:Compactness}
\end{figure}

In Fig.\ref{fig:Compactness}, we show the compactness $C$ of the NTS stars
as a function of the gravitational mass $M_G$ for various breaking scales.
For a fixed breaking scale, $C$ increases monotonically toward the maximum value as $M_G$ increases.
The maximum value of $C$ depends on the breaking scale: $C < 1/3$ for $\eta/M_\text{P} \gtrsim 10^{-1/2}$,
and $C > 1/3$ for $\eta/M_\text{P} \lesssim 10^{-1}$.
In the latter case a NTS star can be so compact that the star has ISCO around it.

\subsection{Binding Energy}
Here,
we consider stability of the NTS stars in terms of the binding energy defined by
$B_G:=M_G - M_{\text{free}}$, where
$M_{\text{free}}$ is sum of mass of free $\psi$ particles that carry totally
the same charge $Q_{\psi}$ of the NTS stars.
A NTS star with $B_G>0$, $(M_G/M_\text{free}>1)$, would disperse into free particles,
i.e., the NTS star is energetically unstable,
while a NTS star with $B_G<0$, $(M_G/M_\text{free}<1)$, is stable against dispersion.

\begin{figure}[H]
\centering
\includegraphics[width=10cm]{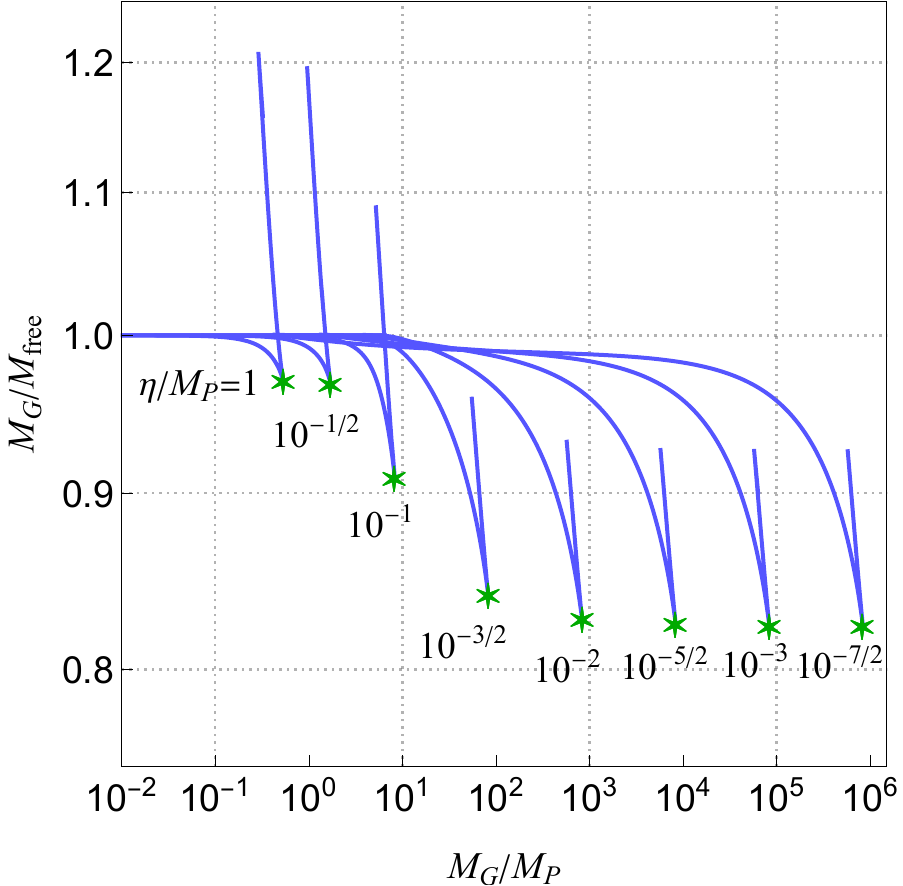}
\caption{
The mass ratio $M_G/M_\text{free}$
as a function of $M_G$
for various breaking scales. The asterisk mark represents the maximum NTS solution
for each breaking scale.
}
\label{fig:Massratio}
\end{figure}

In Fig.\ref{fig:Massratio}, we plot the mass ratio, $M_G/M_\text{free}$, as a function of $M_G$
for various breaking scales. The asterisk marks represent the maximum NTS solutions.
It shows that the maximum NTS stars in all breaking scales have maximum negative binding energies
where $M_G/M_\text{free}<1$,
then solutions near the maximum NTS stars are stable in all breaking scales.

\section{Breaking Scale Dependence}

As shown in the previous section, in each breaking scale,
there exists the maximum NTS star that has maximum gravitational mass.
The mass, the surface radius, and the compactness of the maximum NTS star much depend on
the breaking scale.
Here, we show how these properties of the maximum NTS star depend on the breaking scale.

\begin{figure}[H]
\centering
\includegraphics[width=7cm]{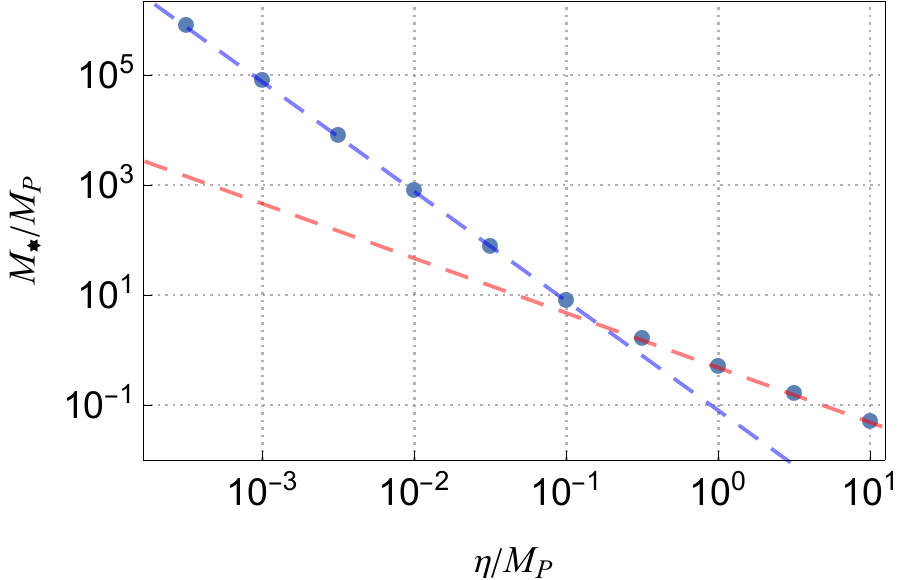}\qquad
\includegraphics[width=7cm]{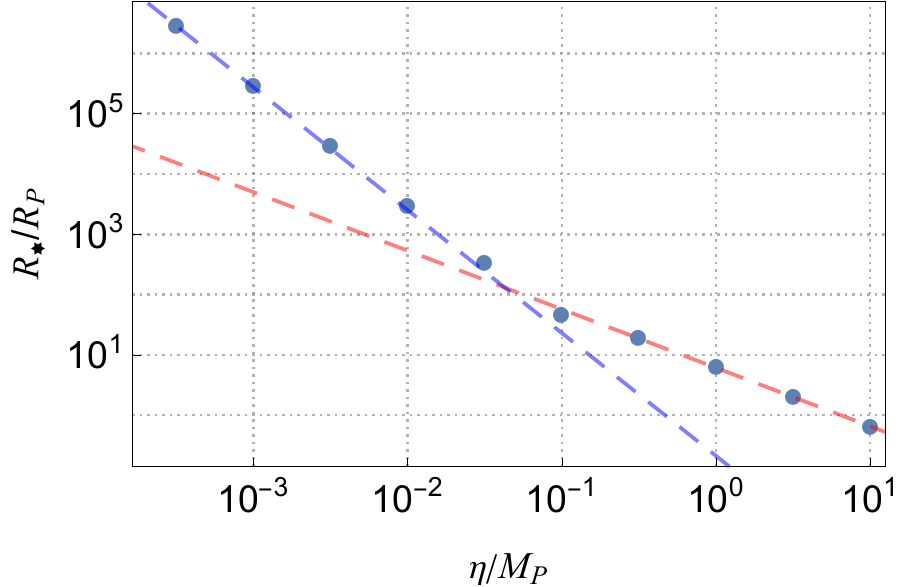}
\caption{
The mass of the maximum NTS stars (left panel),
and the surface radius (right panel)
for each breaking scale $\eta/M_\text{P}$ are plotted as dots.
The plots are fitted by double power laws, respectively.
}
\label{fig:maxMass_scale}
\end{figure}

In Fig.\ref{fig:maxMass_scale}, we plot the mass, $M_*$, and the surface radius, $R_*$,
of the maximum NTS stars for the various values of the breaking scale $\eta/M_\text{P}$.
We observe that the both $M_*$ and $R_*$ obey power laws of $\eta/M_\text{P}$
with two different power indices as
\begin{align}
\frac{M_*}{M_\text{P}}\propto
\begin{cases}
	&\displaystyle \left(\frac{\eta}{M_\text{P}}\right)^{-2} \quad
	\text{for}~ \eta < \eta^M_\text{cr},
\cr
	&\displaystyle \left(\frac{\eta}{M_\text{P}}\right)^{-1} \quad
	\text{for}~ \eta> \eta^M_\text{cr},
\label{eq:Mmax_estimation_smallscale}
\end{cases}
\end{align}
and
\begin{align}
\frac{R_*}{R_\text{P}}\propto
\begin{cases}
	&\displaystyle 	\left(\frac{\eta}{M_\text{P}}\right)^{-2}\quad
	\text{for}~ \eta < \eta^R_\text{cr},
 \\
	&\displaystyle  \left(\frac{\eta}{M_\text{P}}\right)^{-1} \quad
	\text{for}~ \eta > \eta^R_\text{cr},
\end{cases}
\end{align}
where the critical values $\eta^M_\text{cr}$ and $\eta^R_\text{cr}$ are order of $0.1 M_\text{P}$,
and the ratio is $\eta^M_\text{cr}/\eta^R_\text{cr}\sim 3.3$.
The power index in $\eta \gg \eta_\text{cr}$ is the same as mini boson
stars studied in \cite{Kaup:1968zz,Ruffini:1969qy},
and the one in $\eta \ll \eta_\text{cr}$ is
the same as soliton stars studied in \cite{ 
Friedberg:1986tq, Lee:1986tr}.

We consider simple model formulae for $M_*$ and $R_*$ shown in Fig.\ref{fig:maxMass_scale} as
\begin{align}
	M_*	&=\frac{M_\text{cr}}{2}\Big((\eta/\eta^M_\text{cr})^{-2}
 		+(\eta/\eta^M_\text{cr})^{-1}\Big),
\label{M*}
\\
	R_*	&=\frac{R_\text{cr}}{2}\Big((\eta/\eta^R_\text{cr})^{-2}
 		+(\eta/\eta^R_\text{cr})^{-1}\Big),
\label{R*}
\end{align}
where $M_\text{cr}$ and $R_\text{cr}$ are constants.
If we can extrapolate \eqref{M*} and \eqref{R*} for much lower breaking scale than
the cases calculated in the present paper,
typical scales of $M_*$ and $R_*$ for the maximum NTS stars
are listed in Table.\ref{table:estimete_Mmax_R}.
We see that the maximum NTS star would be an astrophysical scale for $\eta\lesssim$ 100GeV.

\begin{table}[htb]
  \begin{tabular}{|c|c|c|c|c|} \hline
     Symmetry Breaking Scale & $M_*[\text{kg}]$ & $R_*[\text{m}]$
     \\ \hline \hline
    $\eta \sim 10^{19}$ GeV & $\mathcal{O}(10^{-8})$ & $\mathcal{O}(10^{-35})$
     \\ \hline
    $\eta \sim 10^{16}$ GeV & $\mathcal{O}(10^{-2})$ & $\mathcal{O}(10^{-29})$
    \\ \hline
    $\eta \sim 10^{2}$ GeV & $\mathcal{O}(10^{26})$ & $\mathcal{O}(10^{-1})$
     \\ \hline
    $\eta \sim 1.0$ GeV & $\mathcal{O}(10^{30})\sim M_{\odot}$ & $\mathcal{O}(10^{3})$
    \\ \hline
   $\eta \sim 1.0$ MeV & $\mathcal{O}(10^{6}M_{\odot})$ & $\mathcal{O}(10^{9})$
     \\ \hline
    $\eta \sim 1.0$ keV & $\mathcal{O}(10^{12}M_{\odot})$ & $\mathcal{O}(10^{15})$
     \\ \hline
    $\eta \sim 1.0$ eV & $\mathcal{O}(10^{18}M_{\odot})$ & $\mathcal{O}(10^{21})$
      \\ \hline
  \end{tabular}
\caption{
The mass, $M_*$, and the radius, $R_*$, of the maximum NTS star for various breaking scales.
The symbol $M_\odot$ denotes the solar mass.
}
  \label{table:estimete_Mmax_R}
\end{table}

According to the model functions \eqref{M*} and \eqref{R*}, we have
\begin{align}
	C_* =\frac{2GM_*}{R_*}
	&=\frac{2G M_\text{cr}}{R_\text{cr}}
		\frac{\eta^M_\text{cr}}{ \eta^R_\text{cr}}
		\frac{( \eta^M_\text{cr}+\eta)}{( \eta^R_\text{cr}+\eta)},
\label{eq:C_estimation_largescale}
\end{align}
then $C_*$ takes the different constant values for $\eta\ll\eta_\text{cr}$ and $\eta\gg\eta_\text{cr}$,
respectively,
and the ratio of them becomes
\begin{align}
	\frac{C_{*}(\eta\ll\eta_\text{cr})}{C_{*}(\eta\gg\eta_\text{cr})}
	=	\frac{\eta^M_\text{cr}} { \eta^R_\text{cr}} \sim 3.3 .
\label{eq:C_value}
\end{align}

\begin{figure}[H]
\centering
\includegraphics[width=8cm]{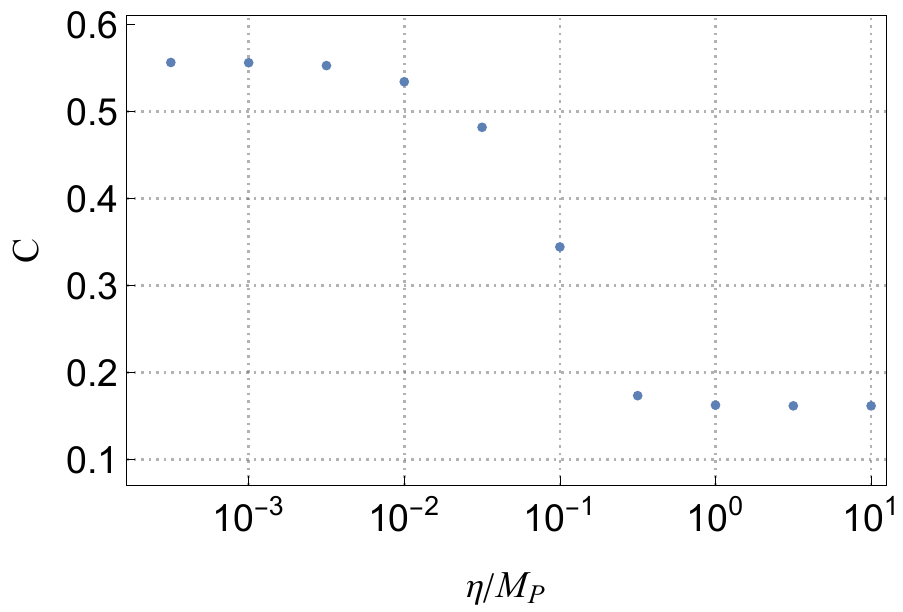}
\caption{
The compactness of the maximum NTS stars for various breaking scales.
}
\label{fig:Compactness_scale}
\end{figure}

In Fig.\ref{fig:Compactness_scale},
we depict the compactness of the maximum NTS stars, $C_*$,
by the use of the numerical values of $M_*$ and $R_*$
as a function of the breaking scale.
From numerical results we see $C_* \sim 0.553$ for $\eta/M_\text{P}\lesssim 10^{-2}$
and $C_* \sim 0.167$ for $\eta/M_\text{P}\gtrsim\eta_\text{cr}$.
Therefore, the maximum NTS stars in $\eta\ll\eta_\text{cr}$ are relativistic self-gravitating
objects that have innermost stable circular orbits while no photon sphere.

\newpage

\section{Conclusions}

We studied the coupled system of field theory that consists of a complex scalar field,
a U(1) gauge field, a complex Higgs scalar field that causes a spontaneous symmetry breaking,
and Einstein gravity.
The system has a dimensionless parameter, $\eta/M_\text{P}$, which represents the ratio of
the symmetry breaking scale to the Plank scale.
We obtained numerical solutions that describe nontopological soliton stars,
parametrized by the angular phase velocity of the complex scalar field, $\Omega$.
The solutions have a variety of properties depending on the parameters $\eta/M_\text{P}$ and $\Omega$.

In the case of the large breaking scale, $\eta/M_\text{P} \gtrsim 0.1$, the solutions are almost determined
by the gravitational field and a scalar field that acquires its mass by the Higgs mechanism.
Then, the solutions are almost same as the mini boson stars obtained in the system of the gravitational
field and a massive complex scalar field.
On the other hand, in the case of small breaking scale, $\eta/M_\text{P} \ll 1$, the solutions are classified
into three types: mini boson stars, matter-interacting NTS stars, and gravitating NTS stars.
For the first type, gravity and a scalar field contribute the solutions as same as the large breaking
scale case. In the second type, interactions between matter fields are important as in the case of
nontopological solitons discussed in Refs.\cite{Ishihara:2018rxg,Ishihara:2019gim,Ishihara:2021iag}.
In the last one, the both matter interaction and self-gravity are important, and NTS star solutions
in this type can have much larger mass than other types.
In the cases of mini boson stars and matter-interacting NTS stars, the gravity is weak
because the lapse function is almost constant everywhere, while in the
case of gravitating NTS stars, gravity requires relativistic description where the lapse varies
significantly.

We found that the maximum mass, which depend on the breaking scale, obeys a double power law:
$M_* \propto \eta^{-1}$ for $\eta \gtrsim \eta_\text{cr}$ and
$M_* \sim \eta^{-2}$ for $\eta \lesssim \eta_\text{cr}$, where $\eta_\text{cr}\sim M_\text{P}/3$ .
If we can extrapolate this for much lower breaking
scale, the maximum mass of the NTS star can be astrophysical scale, the solar mass for $\eta\sim $1GeV
and the claster of galaxies scale for  $\eta\sim $1eV.

We studied the compactness, the gravitational radius over the radius of the NTS stars.
The compactness of NTS stars with maximum mass, $C_*$,
takes the value $C_* \sim 0.167$ for $\eta \gg M_\text{P}$ and $C_*\sim 0.553$ for $\eta \ll M_\text{P}$,
and $C_*$ change in its value quickly around $\eta \sim 0.1 M_\text{P}$.
It means the NTS stars with maximum mass in the lower breaking scale are
relativistically compact object that have the innermost stable circular orbits.
Therefore, the NTS stars in the case $\eta \ll M_\text{P}$ can be seeds of supermassive black holes.

It is an important issue to clarify the stability of the NTS stars.
We would expect that the NTS stars with maximum mass evolve to black holes if they become unstable.
Linear perturbation of the NTS stars would be our next work.

In this paper, we construct NTS star solutions whose internals are filled by kinetic energy
of the scalar fields.
These are self-gravitating solutions of dust balls \cite{Ishihara:2021iag}.
There are other types of NTSs, potential balls and shell balls,
in the model without gravity \cite{Ishihara:2021iag}.
If we take gravity into account, a self-gravitating potential ball would be
a \lq gravastar\rq\ \cite{Mazur:2001fv,Mazur:2004fk}
that join de Sitter and Scwarzshild spacetimes by a spherical shell,
and a self-gravitating shell ball would join Minkowski and Scwarzshild
spacetimes \cite{Kleihaus:2009kr}.
It is also interesting to construct these solutions.

\section*{Acknowledgements}

We would like to thank K.-i. Nakao, H. Yoshino, and M. Minamitsuji
for valuable discussion.
This work was partly supported by Osaka City University Advanced
Mathematical Institute: MEXT Joint Usage/Research Center on Mathematics
and Theoretical Physics JPMXP0619217849.
\newpage

\appendix

\section{Energy-momentum tensor and charge censities}
\label{appendix}

On the assumptions of the scalar and the gauge field forms,
we can reduce the energy-momentum tensor \eqref{eq:T_munu} as
\begin{align}
T^t_t/\eta^4=&-\epsilon
\notag \\
=&-\frac{\left(e^2f^2\alpha^2+(e\alpha-\Omega)^2u^2\right)}{\sigma^2(1-2m/r)}-\left(1-\frac{2m}{r}\right)\left(\left(\frac{df}{dr}\right)^2+\left(\frac{du}{dr}\right)^2\right)
\cr
  &-\frac{\lambda}{4}(f^2-1)^2-\mu f^2u^2-\frac{1}{2\sigma^2}\left(\frac{d\alpha}{dr}\right)^2,
 \label{eq:dimensionless_Ttt}
\\
T^r_r/\eta^4=&p_r
\notag \\
=&\frac{\left(e^2f^2\alpha^2+(e\alpha-\Omega)^2u^2\right)}{\sigma^2(1-2m/r)}+\left(1-\frac{2m}{r}\right)\left(\left(\frac{df}{dr}\right)^2+\left(\frac{du}{dr}\right)^2\right)
\cr
  &-\frac{\lambda}{4}(f^2-1)^2-\mu f^2u^2-\frac{1}{2\sigma^2}\left(\frac{d\alpha}{dr}\right)^2,
 \label{eq:dimensionless_Trr}
 \\
T^{\theta}_{\theta}/\eta^4=&p_{\theta}=T^{\varphi}_{\varphi}/\eta^4=p_{\varphi}
\notag \\
=&
\frac{\left(e^2f^2\alpha^2+(e\alpha-\Omega)^2u^2\right)}{\sigma^2(1-2m/r)}-\left(1-\frac{2m}{r}\right)\left(\left(\frac{df}{dr}\right)^2+\left(\frac{du}{dr}\right)^2\right)
\cr
  &-\frac{\lambda}{4}(f^2-1)^2-\mu f^2u^2+\frac{1}{2\sigma^2}\left(\frac{d\alpha}{dr}\right)^2,
 \label{eq:dimensionless_Tthetatheta}
\end{align}
where $\epsilon$ represents an energy density of the fields, $p_r$ and $p_{\theta}$ denote
pressure in the direction of $r$ and $\theta$.
We have the charge densities as
\begin{align}
  &\rho_{\psi}=-2\sigma^{-2}\left(1-\frac{2m}{r}\right)^{-1}2e(e\alpha-\Omega) u^2,
  \\
  &\rho_{\phi}=-2\sigma^{-2}\left(1-\frac{2m}{r}\right)^{-1}e^2f^2\alpha.
\end{align}

\newpage


\end{document}